\documentclass[prd,aps,nofootinbib,preprintnumbers,showpacs,superscriptaddress,twocolumn,floatfix]{revtex4}

\usepackage{latexsym}
\usepackage{amssymb}
\usepackage{amsfonts}
\usepackage{amsmath}
\usepackage[dvips]{graphicx}

\def\half{\frac{1}{2}}
\def\de{\partial}
\def\lm{{\ell m}}
\def\i{{\rm i}}
\def\l{{\ell}}

\def\l{{\ell }}
\def\cs{C_s^2}
\def\e{\varepsilon}
\def\lam{\Lambda}

\begin{document}
  
  \title{Gravitational waves from pulsations of neutron stars \\
         described by realistic Equations of State}
  
  \author{Sebastiano \surname{Bernuzzi}}
  \affiliation{Dipartimento di Fisica, Universit\`a di Parma, Via G.~Usberti
    7/A, 43100 Parma, Italy}
  \affiliation{INFN, Gruppo Collegato di Parma, Italy}
  \author{Alessandro \surname{Nagar}}
  \affiliation{Institut des Hautes Etudes Scientifiques, 91440 Bures-sur-Yvette,France}
  \affiliation{INFN, Sezione di Torino, Via P.~Giuria 1, Torino, Italy}
  \affiliation{ICRANet, 65122, Pescara, Italy}
  
  \date{\today}
  
  \begin{abstract}
   
    In this work we discuss the time-evolution of nonspherical perturbations of a
    nonrotating neutron star described by a realistic Equation of State (EOS).         
    We analyze 10 different EOS for a large sample of neutron star models.
    Various kind of generic initial data are evolved and the gravitational signals are computed.
    We focus on the dynamical excitation of fluid and spacetime modes 
    and extract the corresponding frequencies.
    We employ a constrained numerical algorithm based on 
    standard finite differencing schemes which permits stable and long term evolutions.
    Our code provides accurate waveforms and allows to capture, 
    via Fourier analysis, the frequencies of the fluid modes with
    an accuracy comparable to that of frequency domain calculations.
    The results we present here are useful for providing comparisons 
    with simulations of nonlinear oscillations of (rotating) neutron 
    star models as well as testbeds for 3D nonlinear codes.

  \end{abstract}
  

  \maketitle

  
  \section{Introduction}
  \label{sec:intrd}  
  
  Neutron stars (NSs) are very compact stars that are born 
  as the result of gravitational collapse~\cite{Shapiro:1983du}.
  They are highly relativistic objects and their internal 
  composition, governed by strong interactions, is, at present, 
  largely unknown. After NS formation (either as the product  
  of gravitational collapse or of the merger of a binary NS system),
  nonisotropic oscillations are typically present. These oscillations
  are damped because of the emission of gravitational waves (GWs).
  In general, the nonspherical oscillations of a NS are characterized by
  two types of proper modes (called {\it quasi-normal} modes, QNMs hereafter):
  fluid modes, which have a Newtonian counterpart, and spacetime 
  (or curvature) modes, which exist only in relativistic stars 
  and are weakly coupled to matter. See Ref.~\cite{lrr-1999-2,Kokkotas:2002ng} for a review.
  The QNMs frequencies carry information about the internal 
  composition of the star and, once detected, they could, 
  in principle, be used to put constraints on the values of 
  mass and radius and thus on the Equation of State (EOS) 
  of a NS~\cite{Andersson:1997rn}. 
  
  In principle, only 3D simulations in full nonlinear General Relativity (GR) 
  with the inclusion of realistic models for the matter composition 
  (as well as electromagnetic fields) can properly investigate 
  the neutron star birth and evolution scenarios. The first successful 
  steps in this direction have been recently done by different 
  groups~\cite{Baiotti:2004wn,shibata:104026,shibata:084021,dimmelmeier:251101}.
  However the complexity of the physical details behind the system 
  and the huge technical/computational costs of these simulations 
  are still not completely accessible, and alternative/approximate 
  approaches to the problem are still meaningful. 
  In particular, we recall the work of Dimmelmeier et al.~\cite{Dimmelmeier:2005zk},
  who simulated oscillating and rotating NSs (described by a 
  polytropic EOS) in the conformally flat (CF) approximation to GR and 
  using specific initial data.
  Another approximate, and historically important, route to studying NS
  oscillations is given by perturbation theory, i.e. by linearizing 
  Einstein's equation around a fixed background (see Ref.~\cite{Kokkotas:2000up}).
  The perturbative approach has proved to be a very reliable 
  method to understand the oscillatory properties of NS as 
  well as a useful tool to calibrate GR nonlinear numerical codes.

  Although most of the work in perturbation theory has been done
  (and is still done) using a frequency-domain approach (in order
  to accurately compute mode frequencies), time-domain simulations
  are also needed to compute full
  waveforms~\cite{Andersson:1996pn,Allen:1997xj,Ferrari:2000sr,Ruoff:2001ux,
  Ruoff:2000et,Ferrari:2003qu,Nagar:2004ns,Nagar:2004av,Nagar:PhD,Ferrari:2004sj,
  Stavridis:2004hg,Passamonti:2005cz,Passamonti:2007tm}.
  In particular, Allen et al.~\cite{Allen:1997xj}, via a multipolar expansion, 
  derived the equations for the even parity perturbations of spherically 
  symmetric relativistic stars and produced explicit waveforms. They 
  argue that, for various kinds of initial data, both fluid and spacetime modes are
  present, but they have different relative amplitudes depending on the
  initial excitation of the system.
  Technically, the problem is reduced to the solution of a set of 3 wave-like 
  hyperbolic equations, coupled to the Hamiltonian constraint: two equations 
  for the metric 
  variables, in the interior and in the exterior of the star, and one equation
  for the fluid variable in the interior. The Hamiltonian constraint is preserved 
  (modulo numerical errors) during the evolution. 
  Ruoff~\cite{Ruoff:2001ux} derived the same set of equations directly from 
  the ADM~\cite{Gravitation:1973} formulation of Einstein's equations and 
  used a similar procedure for their solution. This work showed that
  the presence of spacetime modes in the waveforms strongly depends on the 
  initial data used to initialize the evolution. 
  In particular, conformally flat initial data  can totally suppress the 
  presence of spacetime modes.
  Both studies  use a simplified description of the internal composition 
  of the star, i.e. a polytropic EOS with adiabatic exponent $\Gamma=2$.
  In addition Ref.~\cite{Ruoff:2001ux} explored also the use of one 
  realistic EOS. The author found a numerical instability related to the 
  dip in the sound speed at neutron drip point. This instability was 
  independent either of the formulation of the equations or of the 
  numerical (finite differencing) scheme used. The use of a particular 
  radial coordinate was proposed to cure the problem.

  In this work we reexamine the problem of the evolution of the perturbation
  equations for relativistic stars investigating systematically the gravitational radiation
  emitted from the oscillations of nonrotating neutron star models described
  by a large sample of realistic EOS. 
  We use, specified to the Regge-Wheeler gauge and a static background,
  the general gauge-invariant and coordinate-independent formalism developed  
  in~\cite{Gerlach:1979rw,Gerlach:1980tx,Gundlach:1999bt,MartinGarcia:2000ze}.
  The resulting system of equations is equivalent to the formulation of~\cite{Allen:1997xj,Ruoff:2001ux}.
  For the even-parity perturbation equations, we
  adopt a {\it constrained} numerical
  scheme~\cite{Nagar:PhD,Nagar:2004ns,Nagar:2004av},
  (different from any of those adopted in~\cite{Allen:1997xj,Ruoff:2001ux})
  which permits long-term, accurate and stable evolutions. We use 
  standard Schwarzschild-like coordinate system and we don't need
  the technical complications of Ref.~\cite{Ruoff:2001ux}.
  We evolve various kind of initial data (for odd and even-parity
  perturbations) for 47 neutron star models computed from 10 different EOS.
  We compute and show gravitational waveforms and extract QNMs frequencies. 
  Our accurate results show that there are no relevant
  qualitative differences in the waveforms with respect to previous 
  work limited to polytropic EOS. This was expected since, 
  in first approximation,
  the features of the waves depend only on the star mass and radius 
  (in particular on the compactness).
  The results we report are comprehensive data obtained with a 
  new, efficient numerical code and they complete the information 
  already present in the literature.
  
  The plan of the paper is as follows. 
  In Sec.~\ref{sec:frmls} we briefly review the formalism used 
  and the equations describing the nonspherical perturbations of 
  a spherically symmetric star.
  In Sec.~\ref{sec:emdls} and Sec.~\ref{sec:eos} the construction of the 
  equilibrium star models and the EOS sample are discussed.
  Sec.~\ref{sec:id} deals with the initial data setup, and 
  in Sec.~\ref{sec:res} we present the results.
  We use dimensionless units $c=G=M_\odot=1$, unless otherwise specified 
  for clarity purposes.
  
  
  \section{Perturbation Equations}
  \label{sec:frmls}  
  
  The perturbation equations are obtained by specializing to the nonrotating
  case the general gauge-invariant and coordinate-independent formalism
  for metric perturbations of spherically symmetric spacetimes
  introduced by Gerlach and Sengupta~\cite{Gerlach:1979rw,Gerlach:1980tx} 
  and further developed by Gundlach and Martin-Garcia~\cite{Gundlach:1999bt,MartinGarcia:2000ze}.
  Let us recall that, due to the isotropy of the background spacetime,
  the metric perturbations can be decomposed in multipoles, i.e. expanded
  in tensorial spherical harmonics. These are divided in axial (or
  odd-parity) and polar (or even-parity) modes which, due to the
  spherical symmetry, are decoupled~\footnote{Under a parity transformation
  ($(\theta,\phi)\rightarrow(\pi-\theta,\pi+\phi)$) the axial 
  modes transform as $(-1)^{\l+1}$ and the polar modes as $(-1)^\l$.}.   

  In this work we assume the Regge-Wheeler gauge~\cite{Regge:1957td}.
  The background metric of a static spherical star of radius $R$ 
  and mass $M$, obtained by solving the Tolman-Oppenheimer-Volkoff (TOV) 
  equations~\cite{Gravitation:1973} of hydrostatic equilibrium
  (see below), is written in Schwarzschild-like coordinates as
  \begin{equation}
    g_{\mu\nu}dx^\mu dx^\nu = -e^{2\alpha}dt^2 + e^{2\beta}dr^2 + r^2(d\theta^2 + \sin^2\theta d\phi^2),
  \end{equation}
  where $\alpha(r)$ and $\beta(r)$ are function of $r$ only. The matter 
  is modeled by a perfect fluid:
  \begin{equation}
    T^{\mu\nu} = (\e + p) u^\mu u^\nu + p g^{\mu\nu},
  \end{equation}
  where $p$ is the pressure, $u^\mu$ the fluid 4-velocity,
  and $\e\equiv\rho(1+\epsilon)$ the total energy density.
  Here $\rho$ denotes the rest-mass density and  $\epsilon$ 
  the specific internal energy. The rest-mass density can also 
  be written  in terms of barionic mass $m_B$ and the 
  baryonic number density $n$ as $\rho=m_B n$. 
  The speed of sound is defined as $\cs \equiv \de p/\de\e$.
  The adiabatic exponent is 
  \begin{equation}
    \Gamma \equiv \cs \left(\frac{\e}{p}+1\right).
  \end{equation}  
  In the Regge-Wheeler gauge, the even-parity metric perturbation 
  multipoles are parametrized by three (gauge-invariant) scalar
  functions $(k,\,\chi,\,\psi)$ as~\footnote{We omit hereafter 
  the multipolar indexes for convenience of notation, e.g., $k\equiv k_{\lm}$, 
  $\chi\equiv \chi_{\lm}$ and $\psi=\psi_{\lm}$ }
  \begin{equation}
    \delta g^{(\rm e)}_{\mu\nu} = \left( 
    \begin{array}{cccc}
      (\chi+k)~e^{2\alpha} & -\psi~e^{\alpha+\beta} & 0 & 0 \\
      '' & (\chi+k)~e^{2\beta} & 0 & 0 \\
      '' &  '' & k~r^2 & 0  \\
      '' &  '' & '' & k~r^2\sin^2\theta \\
    \end{array} \right) Y_\lm,
  \end{equation}  
  where $Y_\lm$ are the usual scalar spherical harmonics. 
  Here, $k$ is the perturbed conformal factor, while $\chi$
  is the actual GW degree of freedom. Since the background is
  static, the third function $\psi$ is not independent from
  the others, but can be obtained from $k$ and $\chi$ 
  solving the equation (for $r<R$)~\cite{Gundlach:1999bt}  
  \begin{equation}
  \psi_{,t}=-e^{\alpha-\beta}\left[2e^{2\beta}\left(\dfrac{m}{r^2}+4\pi r p\right)(\chi+k)+\chi_{,r}\right],
  \end{equation}
  where the mass function $m\equiv m(r)$ is 
  defined as $e^{-2\beta(r)} = 1-2m(r)/r$ and represents 
  the mass of the star inside a sphere of radius $r$.
  This equation also holds for $r>R$ with $m(r)=M$.
  
  In addition, when the background is static the metric 
  perturbations are actually described by two degrees 
  of freedom, $(k,\chi)$, only in the interior~\cite{Gundlach:1999bt},
  while only one degree of freedom remains in the exterior. 
  A priori there is no 
  unique way of selecting which evolution equations to use for
  numerical simulations (the ones most convenient mathematically 
  could not be so numerically) so that different formulations of 
  the problem have been numerically explored in the 
  literature~\cite{Allen:1997xj,Ruoff:2001ux,Gundlach:1999bt,Nagar:2004ns}.
  In particular, Ref.~\cite{Nagar:2004ns} showed that it can be useful 
  to formulate the even-parity perturbations problem using a 
  {\it constrained scheme}, with one elliptic and two hyperbolic (wave-like) 
  equations. One  hyperbolic equation is used to evolve $\chi$ in the
  interior and exterior; the other hyperbolic equation serves to
  evolve, in the interior, the perturbation of the relativistic enthalpy
  $H=\delta p/(p+\e)$, where $\delta p$ is the pressure perturbation.
  The system is closed by the elliptic equation, the Hamiltonian constraint,
  that is solved for $k$. Following Ref.~\cite{Nagar:2004ns} we express
  the equations in term of an auxiliary variable $S\equiv\chi/r$, whose
  amplitude tends to a constant for $r\to\infty$ and thus is more convenient
  for the numerical implementation. We recall that the variable $S$ is
  the same used by Ruoff~\cite{Ruoff:2001ux} and the relationship with
  the variables of Allen et al.~\cite{Allen:1997xj} is given by
  $k=F_{\rm Allen}/r$ and $S=e^{-2\alpha}S_{\rm Allen}$. In the star
  interior, $r<R$, the evolution equation for $S$ reads
  \begin{align}
    \label{eq:S}
    -S_{,tt} & + e^{2(\beta-\alpha)}S_{,rr} = e^{2\alpha} \left\{ -\left[ 4\pi r(5p-\e)+\frac{6m}{r^2}\right]S_{,r} \right. \nonumber\\
    & + \left[-4e^{2\beta}\left(\frac{m}{r^2}+4\pi r p\right)^2 - 4\pi(3\e+5p) \right. \nonumber\\
      & \left. + \frac{2}{r^2}\left(1+\frac{m}{r}\right)+\frac{(\ell-1)(\ell+2)}{r^2} \right] S \nonumber\\
    & \left. -2\left[2e^{2\beta} \left(\frac{m}{r^2}+4\pi r p\right)^2 + 8\pi\e
    -\frac{6m}{r^3}\right]\frac{k}{r} \right\},
  \end{align}
  the one for $H$ becomes
  \begin{align}
    \label{eq:H}
    -H_{,tt} &+ \cs e^{2(\beta-\alpha)}H_{,rr} = 
    e^{2\alpha}\left\{ \left[\frac{m}{r^2}(1+\cs) \right.\right.\\
      & \left. + 4\pi pr\left(1-2\cs\right) + \left(4\pi r\e -\frac{2}{r}\right)\cs \right]H_{,r} \nonumber\\ 
    & - \left[4\pi(p+\e)(3\cs+1) - \cs\frac{\lam}{r^2}\right]H \nonumber\\ 
    & +\frac{1}{2}\left(\frac{m}{r^2}+4\pi pr\right)(1-\cs)(rS_{,r}-k_{,r}) \nonumber\\
    & + \left[\frac{2(m+4\pi p r^3)^2}{r^3(r-2m)} -4\pi \cs(3p +\e) \right](rS + k)\nonumber
  \end{align}
  and finally the Hamiltonian constraint is 
  \begin{align}
    \label{eq:Ham}
    \left(1-\frac{2m}{r}\right)k_{,rr} & + \left[\frac{2}{r} - \frac{3m}{r^2}-4\pi \e
      r\right]k_{,r} - \left[\frac{\lam}{r^2}-8\pi \e\right]k = \nonumber\\ 
    & - \frac{8\pi ( p+\e)}{\cs} H + \left(1-\frac{2m}{r}\right)S_{,r} \nonumber\\ 
    & +\left[\frac{2}{r}-\frac{2m}{r^2}+\frac{\lam}{2r} -8\pi \e r\right]S,
  \end{align}
  where $\Lambda\equiv\ell(\ell+1)$.  Eqs.~\eqref{eq:S} and~\eqref{eq:Ham} are
  also valid in the exterior, with $\cs=p=H=0$, $m(r)=M$ and $e^{2\alpha}=1-2M/r$.
  Since in the star exterior the spacetime is described by the Schwarzschild 
  metric, the perturbation equations can be combined together in 
  the Zerilli equation~\cite{Zerilli:1970se}
  \begin{equation}
    \label{eq:ZMequation}
    \Psi^{(\rm e)}_{,tt}-\Psi^{(\rm e)}_{,r_*r_*} + V^{(\rm e)}_\l\Psi^{(\rm e)} = 0
  \end{equation}  
  for a single, gauge-invariant, master function $\Psi^{(\rm e)}$, the 
  Zerilli-Moncrief function~\cite{Zerilli:1970se,Moncrief:1974am}. 
  The function $V^{(\rm e)}_\l$ is the Zerilli potential 
  (see for example~\cite{Nagar:2005ea}) and $r_*=r + 2M \ln[r/(2M) -1]$ 
  is the Regge-Wheeler tortoise coordinate. 
  In terms of the gauge-invariant functions $\chi$ and $k$, $\Psi^{(\rm e)}$
  reads
  \begin{align}
    \label{eq:ZM}
    \Psi^{(\rm e)} = \dfrac{2r(r-2M)}{\Lambda[(\Lambda -2)r +6M]}\left[\chi-
      r k_{,r} + \dfrac{r\Lambda+2M}{2(r-2M)} k\right].
  \end{align}
  The inverse equations can be found, for instance, in
  Ref.~\cite{Ruoff:2001ux}. In our notation they read
  \begin{align}
    \label{eq:k}
    k &= 2e^{2\alpha}\Psi^{(\rm e)}_{,r} + \left\{ \dfrac{\lam}{r}-\frac{12M e^{2\alpha}}{r[r(\lam -2)+6M]}\right\}\Psi^{(\rm e)},\\
     \label{eq:chi}
    \chi&= 2e^{2\alpha}\Psi^{(\rm e)}_{,rr} + \frac{2M}{r^2}\left(1-\frac{6 r e^{2\alpha}}{r(\lam -2)+6M} \right)\Psi^{(\rm e)}_{,r}\nonumber \\
    & + \frac{2}{r^2}\left[\frac{3M}{r}-\lam+\frac{6M}{r(\lam -2)+6M}\left( 3-\frac{8M}{r}\right.\right. \nonumber \\
      & -\left.\left. \frac{6M e^{2\alpha}}{r(\lam -2)+6M}\right)\right]\Psi^{(\rm e)}.
  \end{align}
  Let's mention briefly the boundary conditions to impose to these equations.
  At the center of the star all the function must be regular, and this leads to
  the conditions:
  \begin{align}
    \chi &\sim r^{\ell+2},\label{eq:BC1}\\
    k &\sim r^{\ell+1}, \label{eq:BC2}\\
    H &\sim r^\ell. \label{eq:BC3} 
  \end{align}
  At the star surface $S$ is continuous as well as its first and second 
  radial derivatives. On the contrary, $k$ and its first radial derivative 
  are continuos but $k_{,rr}$ can have a discontinuity due the term 
  $8\pi( p+\e)H/\cs$ in Eq.~\eqref{eq:Ham}.
  At the star surface, $r=R$, Eq.~\eqref{eq:H} reduces to an ODE 
  for $H$, that is solved accordingly.
  
  On a static background, the odd-parity perturbations are described
  by a single, gauge-invariant, dynamical variable $\Psi^{(\rm o)}$,   
  that is totally decoupled from matter. This function satisfies a 
  wave-like equation of the form~\cite{Gerlach:1979rw,Chandrasekhar:1991fi}
  \begin{equation}
    \label{eq:RWequation}
    \Psi^{(\rm o)}_{,tt}-\Psi^{(\rm o)}_{,\bar{r}_*\bar{r}_*} + V^{(\rm
    o)}_{\ell}\Psi^{(\rm o)} = 0 ,
  \end{equation} 
  with a potential
 \begin{equation}
    \label{eq:RWpotential}
    V^{(\rm o)}_{\l} = e^{2\alpha}\left(\frac{6m}{r^3} + 4\pi(p -\e)-\frac{\lam}{r^2} \right).
 \end{equation}
  This equation has been conveniently written in terms of the
  ``star-tortoise'' coordinate $\bar{r}_*$ defined as $\partial \bar{r}_*/\partial r = e^{\beta-\alpha}$.
 In the exterior, $\bar{r}_*$ reduces to the Regge-Wheeler tortoise
 coordinate $r_*$ introduced above and Eq.~\eqref{eq:RWequation} 
 becomes the well-known Regge-Wheeler equation~\cite{Regge:1957td}.
 The relation between $\Psi^{(\rm o)}$ and the odd-parity metric multipoles
 is given, for example, by Eqs.~(19)-(20) of Ref.~\cite{Nagar:2005ea}.
  
 The principal quantities we want to obtain are the gauge-invariant 
 functions  $\Psi^{(\rm e/o)}$. These functions are directly 
 related to the ``plus'' and ``cross'' polarization amplitudes 
 of the GWs by (see e.g.~\cite{Nagar:2005ea,Martel:2005ir}):
  \begin{equation}
    \label{eq:h+hp}
    h_+ - \i h_\times = \frac{1}{r}\sum_{\l=2}^{\infty}\sum_{m=-\l}^{\l} N_\l
    \left( \Psi^{(\rm e)}_{\lm} + \i \Psi^{\rm (o)}_{\lm} \right)
    {}_{-2}Y_{\lm}(\theta,\phi), 
  \end{equation}
  where $N_\l=\sqrt{(\l+2)(\l+1)\l(\l-1)}$ and ${}_{-2}Y_{\l m}$ are the
  spin-weighted spherical harmonics of spin-weight $s=-2$. The GWs luminosity 
  at infinity is given by
  \begin{equation}
    \label{eq:GWenergy}
    \frac{dE}{dt} = \frac{1}{16\pi}\sum_{\l=2}^{\infty}\sum_{m=-\l}^{\l} N_{\l}^2\left(
    \left|\dot{\Psi}^{(\rm e)}_{\lm}\right|^2 + \left|\dot{\Psi}^{(\rm o)}_{\lm}\right|^2\right), 
  \end{equation}
  where the overdot stands for derivative with respect to coordinate time $t$.
  The energy spectrum reads
  \begin{equation}
    \frac{dE}{d\omega} = \frac{1}{16\pi^2}
    \sum_{\l=2}^{\infty}\sum_{m=-\l}^{\l}~N_\l^2\omega^2  \left(
    \left|\tilde{\Psi}^{(\rm e)}_{\lm}\right|^2 + \left|\tilde{\Psi}^{(\rm o)}_{\lm}\right|^2,
    \right) 
  \end{equation}
  where $\tilde{\Psi}_{\l m}^{(\rm e/o)}$ indicates the Fourier transform of
  $\Psi^{(\rm e/o)}_{\l m}$, $\omega=2\pi\nu$ and $\nu$ is the frequency.
  
  
  \section{Equilibrium Stellar Models}
  \label{sec:emdls}  
  
  The equilibrium configuration of a spherically symmetric and relativistic star 
  is the solution of the TOV equations
  \begin{eqnarray}
    m_{, r} &=& 4\pi r^2 \e \;\;\;, \nonumber\\
    p_{, r} &=& -(\e+p)\alpha_{, r} \;\;\;,\label{eq:TOV} \\
    \alpha_{, r} &=& \half\frac{(m+4\pi r^3 p)}{(r^2 -2mr)} \;\;\;, \nonumber
  \end{eqnarray}
  with the boundary conditions 
  \begin{eqnarray}
    \label{eq:TOVbm}
    m(0) &=& 0\\
    \label{eq:TOVbp}
    p(R) &=& 0 \\ 
    \label{eq:TOVba}
    \alpha(R) &=& \ln\left(1-\frac{2M}{R}\right). 
  \end{eqnarray}
  The system is closed with an EOS $p(\rho)$. 
  Eq.~(\ref{eq:TOVbp}) formally defines the star radius, $R$. 
  Eq.~(\ref{eq:TOV}) define the structure of the fluid 
  and its spacetime in the interior, i.e. $r<R$, then the solution is matched
  at the exterior Schwarzschild solution, Eq.~(\ref{eq:TOVba}).
  
  
  \section{Equations of State}
  \label{sec:eos}  
   
  \begin{table}[t]
    \caption{\label{tab:eos} A list of the EOS name and references that we use in this work.}
    \begin{ruledtabular}
      \begin{tabular}{ccc}
	Name & Authors & References \\
	\hline
	A & Pandharipande & \cite{Pandharipande:1971:eosA}\\
	B & Pandharipande & \cite{Pandharipande:1971:eosB}\\
	C & Bethe and Johnson & \cite{Bethe:1974:eosC}\\
	FPS & Lorenz, Ravenhall and Pethick & \cite{Friedman:1981:FP,Lorenz1993:FPS}\\
	G & Canuto and Chitre & \cite{Canuto:1974:eosG}\\
	L & Pandharipande and Smith & \cite{Arnett:1976dh}\\
	N & Walecka and Serot & \cite{Walecka:1974:eosN}\\
	O & Bowers, Gleeson and Pedigo & \cite{Bowers:1975:eosO,Bowers:1975:eosOb}\\
	SLy & Douchin and Haensel & \cite{Douchin:2001sv}\\
	WFF & Wiringa, Fiks and Farbroncini & \cite{Wiringa:1988tp}\\
      \end{tabular}
    \end{ruledtabular}
  \end{table}

  Neutron stars are composed by high density baryonic matter. 
  The exact nature of the internal structure, determined essential 
  by strong interactions, is unknown. To model the neutron star interior 
  (approximated) many-body theories with effective Hamiltonians 
  are usually employed. The principal assumptions are that the 
  matter is strongly degenerate and that it is at the thermodynamics  
  equilibrium. Consequently, temperature effects can be neglected 
  and the matter is in its ground state (cold catalyzed matter). 
  Under this conditions the EOS has one-parameter character:
  $\e(n)$ and $p(n)$ or $\e(p)$.
  
  \begin{figure}[t]
    \begin{center}
      \includegraphics[width=0.5\textwidth]{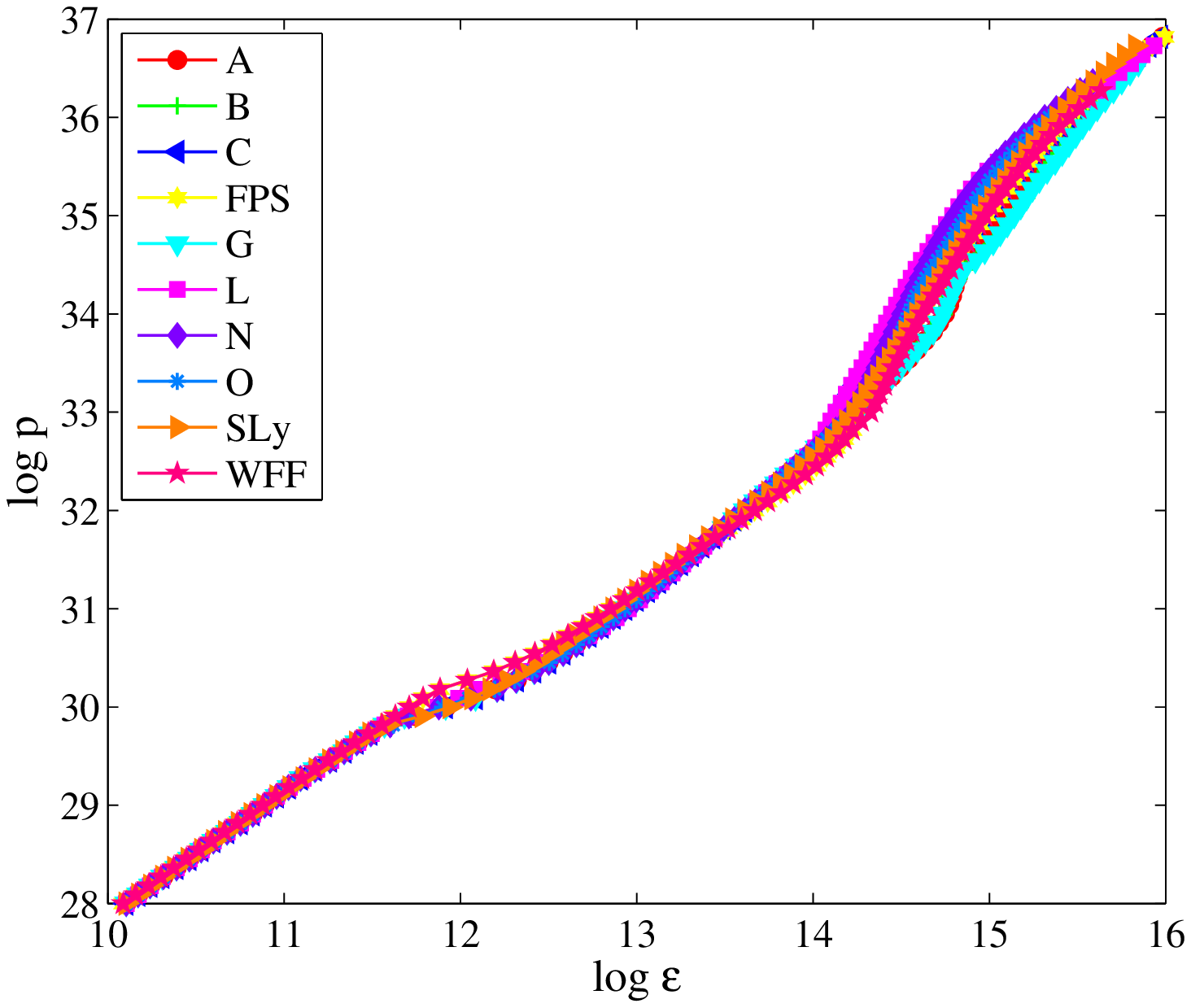}\\
      \includegraphics[width=0.5\textwidth]{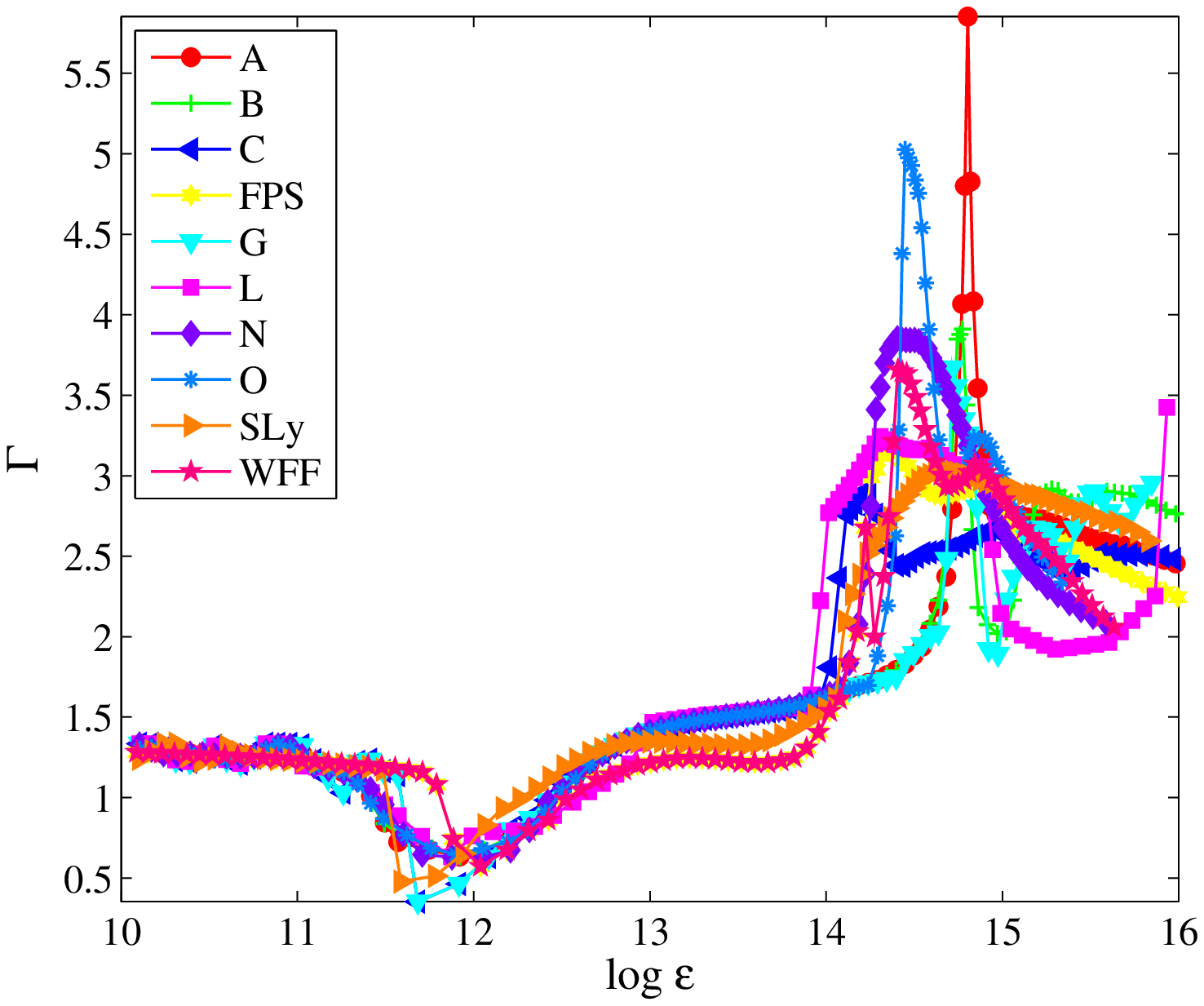}\\
      \caption{\label{fig:eos} Pressure (top) and the adiabatic exponent (bottom)
        as a function of the total energy density for various EOS.
        Notice here we are using cgs units.}
    \end{center}
  \end{figure}
  
  The composition of a neutron star consists qualitatively of three parts
  separated by transition points, see Fig.~\ref{fig:eos}. 
  At densities below the neutron drip, 
  $\e<\e_{\rm d}\sim10^{11}$ $\rm gcm^{-3}$ (the \emph{outer crust}) 
  the nuclei are immersed in an electron gas and the electron pressure 
  is the principal contribute to the EOS. In the \emph{inner crust}, 
  $\e_{\rm d}<\e<10^{14}$ $\rm gcm^{-3}$,
  the gas is also composed of a fraction of neutrons unbounded from the nuclei and the 
  EOS softens due to the attractive long-range behaviour of the strong interactions.
  For $\e>10^{14}$ $\rm gcm^{-3}$ 
  a homogeneous plasma of nucleons, electrons, muons and 
  other baryonic matter (e.g. hyperons), composes the \emph{core}
  of the star. In this region the EOS stiffens because of the repulsive 
  short-range character of the strong interactions. 
  The bottom panel of Fig.~\ref{fig:eos} shows the adiabatic exponent 
  which, under the assumption of thermodynamics equilibrium, determines 
  the response of pressure to a local perturbation of density.
  We mention that, as explained in detail in Ref~\cite{Douchin:2001sv},
  in a pulsating NS the ``actual adiabatic exponent'' can be higher than that 
  obtained from the EOS because the timescale of the beta processes are longer 
  than the dynamical timescales of the pulsations. Thus the fraction of particles 
  in a perturbed fluid element is assumed 
  fixed to the unperturbed values (frozen composition).
  We refer the reader to~\cite{Haensel:2007yy} for all the details 
  on neutron star structure and the complex physics behind it 
  (e.g. elasticity of the crust, possible superfluid interior core, magnetic fields).

  \begin{table*}[H]
    \caption{\label{tab:equil} Neutron Star models. From left to right the
      columns report: the name of the model, the EOS type, the mass $M$, the 
      radius $R$, the compactness $M/R$, the central energy density $\e_{\rm c}$
      and the central pressure $p_c$.}
    \begin{ruledtabular}
      \begin{tabular}{ccccccc}
	Model & EOS  & $M$ & $R$ & $M/R$ & $\e_{\rm c}$ & $p_{\rm c}$ \\
	\hline
	A10 & A & 1.00 & 6.55 & 0.15 & 1.96$\times10^{-3}$ & 2.35$\times10^{-4}$\\
	A12 & A & 1.20 & 6.51 & 0.18 & 2.38$\times10^{-3}$ & 3.76$\times10^{-4}$\\
	A14 & A & 1.40 & 6.39 & 0.22 & 3.01$\times10^{-3}$ & 6.50$\times10^{-4}$\\
	A16 & A & 1.60 & 6.04 & 0.26 & 4.46$\times10^{-3}$ & 1.49$\times10^{-3}$\\
	Amx & A & 1.65 & 5.60 & 0.29 & 6.78$\times10^{-3}$ & 3.24$\times10^{-3}$\\
	B10 & B & 1.00 & 5.78 & 0.17 & 1.37$\times10^{-3}$ & 5.13$\times10^{-4}$\\
	B12 & B & 1.20 & 5.53 & 0.22 & 1.59$\times10^{-3}$ & 1.02$\times10^{-3}$\\
	B14 & B & 1.40 & 4.96 & 0.28 & 1.88$\times10^{-3}$ & 3.42$\times10^{-3}$\\
	Bmx & B & 1.41 & 4.73 & 0.30 & 4.65$\times10^{-3}$ & 5.33$\times10^{-3}$\\
	C10 & C & 1.00 & 8.15 & 0.12 & 3.35$\times10^{-3}$ & 1.05$\times10^{-4}$\\
	C12 & C & 1.20 & 8.05 & 0.15 & 4.45$\times10^{-3}$ & 1.67$\times10^{-4}$\\
	C14 & C & 1.40 & 7.90 & 0.18 & 7.76$\times10^{-3}$ & 2.65$\times10^{-4}$\\
	C16 & C & 1.60 & 7.67 & 0.21 & 9.76$\times10^{-3}$ & 4.49$\times10^{-4}$\\
	Cmx & C & 1.85 & 6.69 & 0.28 & 1.18$\times10^{-3}$ & 1.94$\times10^{-3}$\\
	FPS10 & FPS & 1.00 & 7.32 & 0.14 & 1.44$\times10^{-3}$ & 1.50$\times10^{-4}$\\
	FPS12 & FPS & 1.20 & 7.30 & 0.16 & 1.75$\times10^{-3}$ & 2.30$\times10^{-4}$\\
	FPS14 & FPS & 1.40 & 7.22 & 0.19 & 2.22$\times10^{-3}$ & 3.64$\times10^{-4}$\\
	FPS16 & FPS & 1.60 & 7.02 & 0.23 & 4.77$\times10^{-3}$ & 6.38$\times10^{-4}$\\
	FPSmx & FPS & 1.80 & 6.22 & 0.29 & 1.45$\times10^{-3}$ & 2.50$\times10^{-3}$\\
	G10 & G & 1.01 & 5.76 & 0.17 & 1.73$\times10^{-3}$ & 5.36$\times10^{-4}$\\
	G12 & G & 1.20 & 5.42 & 0.22 & 2.10$\times10^{-3}$ & 1.20$\times10^{-3}$\\
	Gmx & G & 1.36 & 4.62 & 0.29 & 2.72$\times10^{-3}$ & 5.65$\times10^{-3}$\\
	L10 & L & 1.00 & 9.61 & 0.10 & 5.55$\times10^{-3}$ & 4.15$\times10^{-5}$\\
	L12 & L & 1.20 & 9.76 & 0.12 & 3.52$\times10^{-3}$ & 5.60$\times10^{-5}$\\
	L14 & L & 1.40 & 9.89 & 0.14 & 5.03$\times10^{-3}$ & 7.41$\times10^{-5}$\\
	L16 & L & 1.60 & 9.98 & 0.16 & 7.81$\times10^{-4}$ & 9.75$\times10^{-5}$\\
	Lmx & L & 2.68 & 9.23 & 0.29 & 5.80$\times10^{-4}$ & 9.17$\times10^{-4}$\\
	N10 & N & 1.00 & 8.81 & 0.11 & 6.38$\times10^{-4}$ & 5.70$\times10^{-5}$\\
	N12 & N & 1.20 & 8.98 & 0.13 & 7.05$\times10^{-4}$ & 7.59$\times10^{-5}$\\
	N14 & N & 1.40 & 9.13 & 0.15 & 7.81$\times10^{-4}$ & 9.97$\times10^{-5}$\\
	N16 & N & 1.60 & 9.24 & 0.17 & 2.33$\times10^{-3}$ & 1.30$\times10^{-4}$\\
	Nmx & N & 2.63 & 8.65 & 0.30 & 7.08$\times10^{-4}$ & 1.20$\times10^{-3}$\\
	O10 & O & 1.00 & 8.29 & 0.12 & 1.41$\times10^{-3}$ & 7.39$\times10^{-5}$\\
	O12 & O & 1.20 & 8.41 & 0.14 & 1.64$\times10^{-3}$ & 1.03$\times10^{-4}$\\
	O14 & O & 1.40 & 8.50 & 0.16 & 1.96$\times10^{-3}$ & 1.43$\times10^{-4}$\\
	O16 & O & 1.60 & 8.54 & 0.19 & 2.45$\times10^{-3}$ & 1.97$\times10^{-4}$\\
	Omx & O & 2.38 & 7.75 & 0.31 & 5.15$\times10^{-3}$ & 1.66$\times10^{-3}$\\
	SLy10 & SLy4 & 1.00 & 7.78 & 0.13 & 7.72$\times10^{-4}$ & 1.12$\times10^{-4}$\\
  	SLy12 & SLy4 & 1.20 & 7.80 & 0.15 & 8.41$\times10^{-4}$ & 1.66$\times10^{-4}$\\
	SLy14 & SLy4 & 1.40 & 7.78 & 0.18 & 9.17$\times10^{-4}$ & 2.45$\times10^{-4}$\\
	SLy16 & SLy4 & 1.60 & 7.70 & 0.21 & 2.57$\times10^{-3}$ & 3.70$\times10^{-4}$\\
	SLymx & SLy4 & 2.05 & 6.71 & 0.30 & 8.74$\times10^{-4}$ & 2.50$\times10^{-3}$\\
	WFF10 & WFF3 & 1.00 & 7.29 & 0.14 & 9.86$\times10^{-4}$ & 1.46$\times10^{-4}$\\
	WFF12 & WFF3 & 1.20 & 7.30 & 0.16 & 1.10$\times10^{-3}$ & 2.18$\times10^{-4}$\\
	WFF14 & WFF3 & 1.40 & 7.26 & 0.19 & 1.23$\times10^{-3}$ & 3.34$\times10^{-4}$\\
	WFF16 & WFF3 & 1.60 & 7.14 & 0.22 & 3.36$\times10^{-3}$ & 5.48$\times10^{-4}$\\
	WFFmx & WFF3 & 1.84 & 6.40 & 0.29 & 1.18$\times10^{-3}$ & 2.18$\times10^{-3}$
      \end{tabular}
    \end{ruledtabular}
  \end{table*}
  

  \begin{figure}[t]
    \begin{center}
      \includegraphics[width=0.5\textwidth]{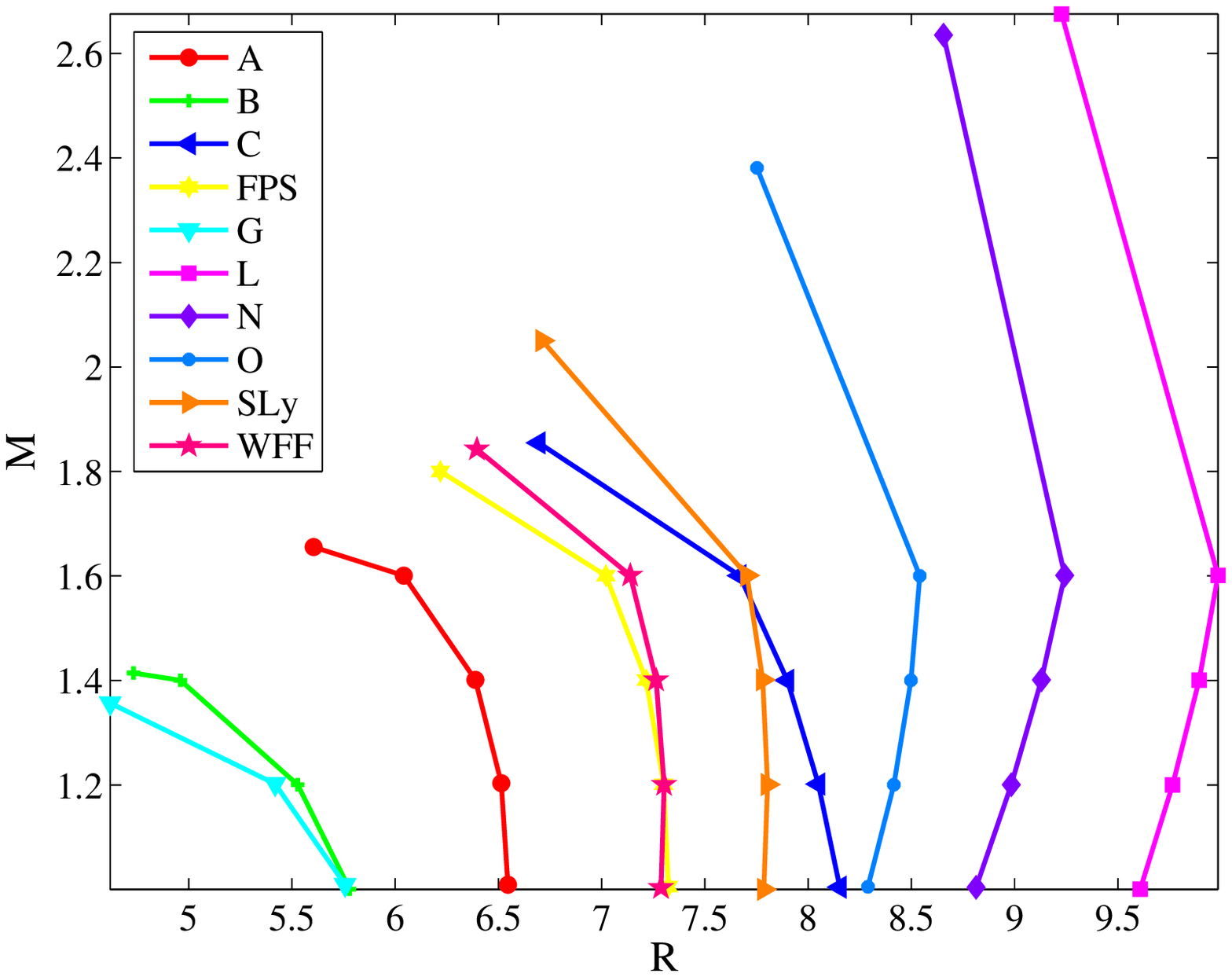}
      \includegraphics[width=0.5\textwidth]{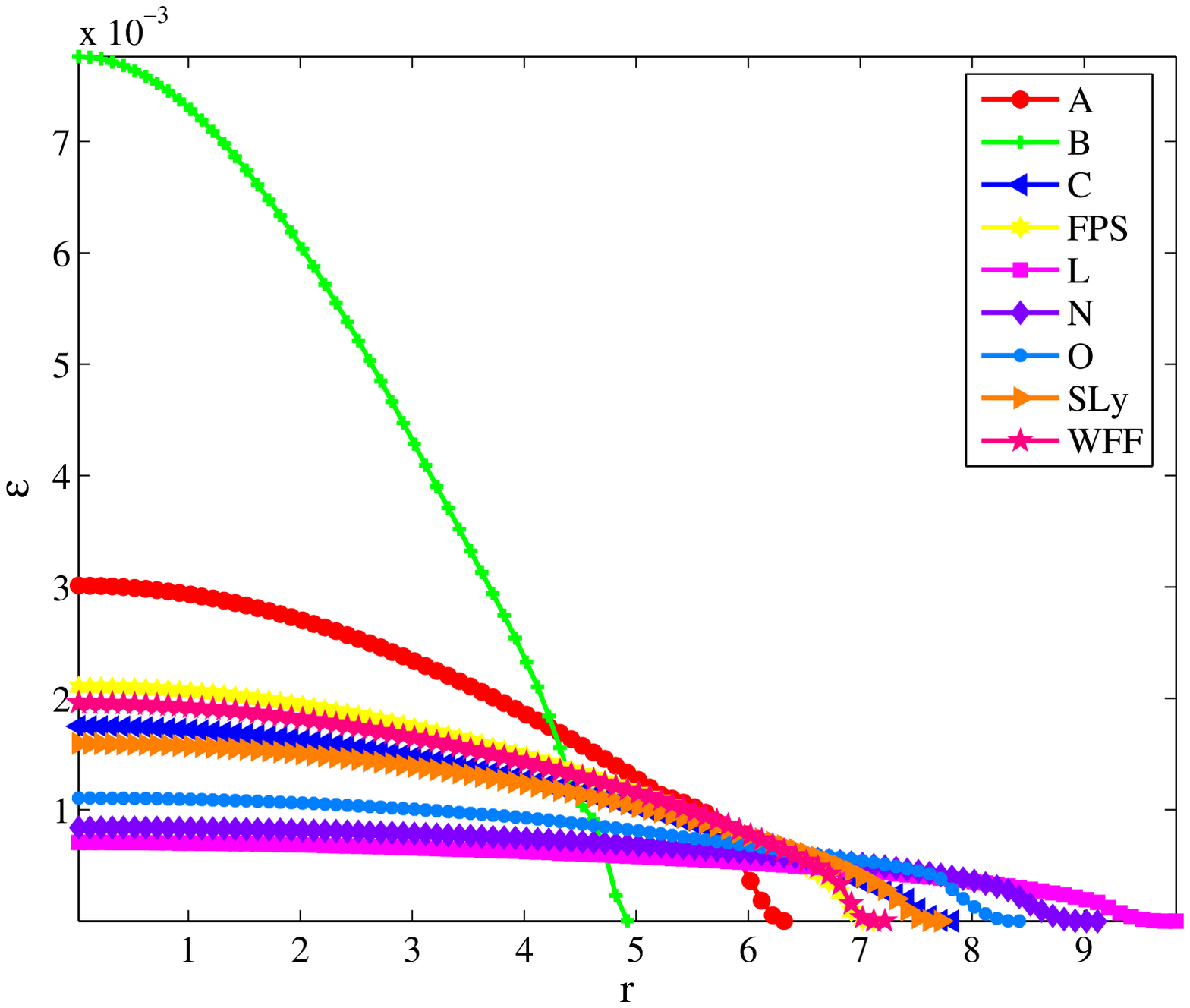}
      \caption{\label{fig:equil} Top panel: mass versus radius for the NS  
	models of Table~\ref{tab:equil}. Bottom panel: profile of the 
	total energy density $\e$ for all models with $M=1.4$.}
    \end{center}
  \end{figure}
  
  Most of the mass (see the bottom panel of Fig.~\ref{fig:equil}) is 
  constituted by high density matter, so that the maximum mass of the 
  star is essentially determined by the EOS of the core. On the other hand,
  the star radius strongly depends on the properties of the matter 
  at low densities, due to Eq.~(\ref{eq:TOVbp}).
  
  To compensate the ignorance on the interior part of the star
  it is common to consider a large set of EOS derived from different models. 
  In our work we employ 7 realistic EOS already used in \cite{Arnett:1976dh},
  and in many other works (see for example 
  Refs.~\cite{1983ApJS...53...73L,Andersson:1997rn,Benhar:2004xg,Kokkotas:2000up,Benhar:1998au} 
  on pulsations of relativistic stars, and
  \cite{Salgado:1994:sbghI,Stergioulas:1994ea,Nozawa:1998ak} on 
  equilibrium models of rotating stars). 
  Maintaining the same notation of~\cite{Arnett:1976dh}, they are called 
  A, B, C, G, L, N and O EOS. Most of the models in the sample are based 
  on non-relativistic interactions modeled with Reid soft core type potentials. 
  EOS N \cite{Walecka:1974:eosN} and O~\cite{Bowers:1975:eosO,Bowers:1975:eosOb}
  are instead based on relativistic interaction and many-body theories. 
  Model  G~\cite{Canuto:1974:eosG} is an extremely soft EOS, while L~\cite{Arnett:1976dh}
  is extremely stiff. EOS A~\cite{Pandharipande:1971:eosA} and
  C~\cite{Bethe:1974:eosC} are of intermediate stiffness. 
  In addition, we use the FPS EOS~\cite{Friedman:1981:FP,Lorenz1993:FPS}, 
  and the SLy EOS~\cite{Douchin:2001sv}, modeled by Skyrme effective interactions.
  The FPS EOS, in particular is a modern version of Friedman and Pandharipande 
  EOS~\cite{Friedman:1981:FP}.
  The last EOS considered is the UV14+TNI (here renamed WFF) 
  EOS of~\cite{Wiringa:1988tp}, which is
  an intermediate stiffness EOS based on two-body Urbana 
  UV14 potential 
  with the phenomenological three-nucleon TNI interaction. 
  The composition is assumed to be of neutrons.
  For all the EOS models the inner crust is described by the 
  BBS \cite{Baym:1971:BBP} or the HP94 \cite{Haensel:1994:HP94} EOS, 
  while for outer crust the BPS EOS \cite{Baym:1971:BPS} is used.
  We refer to Table~\ref{tab:eos} and cited references for further details.
    
  Realistic EOS are usually given through tables. To use them in a numerical
  context it is necessary to interpolate between the tabulated values.
  The interpolation can not be chosen arbitrarily
  but must properly take into account the First Law of Thermodynamics,
  see e.g.~\cite{Haensel:1982},
  that, in the case of a temperature independent EOS, reads:
  \begin{equation}
    \label{eq:1lawther}
    p(n)=n^2\frac{d}{dn}\left(\frac{\e}{n}\right)
  \end{equation}
  A thermodynamically consistent procedure is described in~\cite{Swesty:1996} 
  and it is based on Hermite poynomials. Essentially the method permits
  to interpolate a function forcing the match on the tabulated points 
  both of the function and of its derivatives.
  We implement this scheme using 
  cubic Hermite poynomials as already done in \cite{Nozawa:1998ak}, 
  The procedure used is described in detail in Appendix~\ref{app:int}.

  \section{Initial Data}
  \label{sec:id}  
  
  In principle the choice of the initial data for the perturbation equations 
  should take into account, at least approximately, the astrophysical scenario
  in which the neutron star is born. Such scenario could be, for example 
  the gravitational collapse or the merger of two neutron stars.
  Only long-term simulations in full general relativity can investigate 
  highly nonlinear and nonisotropic system until they 
  settle down in a, almost spherical, quasi-equilibrium configuration.
  A perturbative analysis, like the one we propose, could then start 
  from this point once the metric and matter tensors had been projected 
  along the corresponding (tensorial) spherical harmonics.
  This approach, in principle possible, is however beyond the scopes of the present work.
  
  Inspired by previous perturbative calculations~\cite{Allen:1997xj,Ruoff:2001ux},
  we consider different kinds of initial data such that they are the simplest, 
  well-posed and involve perturbations of both the fluid and/or the metric
  quantities.
  
  In the case of the even parity perturbations
  we start the evolutions from 3 different initial excitations 
  of fluid and matter variables:
  \begin{enumerate}
  \item \emph{Conformally Flat Initial Data.} 
    We set $S(0,r)=0$ and give a fluid perturbation of type:
    \begin{equation}
      \label{eq:id:H}
      H(0,r) = A \left(\frac{r}{R}\right)^{\ell-1}\sin\left(\pi (n+1) \frac{r}{R}\right).
    \end{equation}
    The function $k(0,r)$ is computed consistently solving the Hamiltonian
    constraint. The profile of $H$ in Eq.~\eqref{eq:id:H} is chosen in 
    order to approximate the behavior of an enthalpy 
    eigenfunction with $n$ nodes. In this way, only some modes can be
    (prominently) excited.
    Since we would like to focus on the principal fluid modes 
    we chose a \emph{zero-nodes} initial data setting $n=0$.
  \item \emph{Radiative Initial Data.}
    We set $k(0,r)=0$ and $H(0,r)$ as Eq.~\eqref{eq:id:H}. 
    The function $S(0,r)$ is computed consistently solving 
    the Hamiltonian constraint.
  \item \emph{Scattering-like Initial Data.}
    We set $H(0,r)=0$ and $\Psi^{(\rm e)}(0,r)$ as a Gaussian pulse:
    \begin{equation}
      \label{eq:id:gauss}
      \Psi^{(\rm e)}(0,r) = A\exp{\left(-\dfrac{(r-r_{\rm c})^2}{b^2}\right)},
    \end{equation}
    with $r_{\rm c}=70M$ and $b=M$.
    The functions $k(0,r)$ and $S(0,r)$ are computed consistently 
    from Eqs.~\eqref{eq:k}-\eqref{eq:chi}.
  \end{enumerate}
  Initial data of type 1 and 2 are choosen to be time-symmetric 
  ($S_{,t}=k_{,t}=H_{,t}=0$). 
  On the one hand, this choice can be physically questionable
  because the system has an unspecified amount of 
  incoming radiation in the past.
  On the other hand, it is the simplest choice and guaratees that 
  the momentum constraints are trivially satisfied and only the 
  Hamiltonian constraint needs to be used for the setup.
  The Gaussian in type 3 initial data is \emph{ingoing}, i.e. 
  $\Psi_{,t}=\Psi_{,r_{*}}$, and the derivatives of the 
  other variables are computed consistently with this choice.
  
  In the case of odd-parity perturbations we start the evolution using 
  an ingoing narrow gaussian as in Eq.~\eqref{eq:id:gauss} for 
  $\Psi^{(\rm o)}(0,r)$.
  The amplitude of the perturbation A is everywhere chosen equal to 0.01.
  
  
  \section{Results}
  \label{sec:res}  
  
  For each EOS, we study a rapresentative set of models with 
  $M=1$, $1.2$, $1.4$, $1.6$ and a model whose mass is close 
  to the maximum mass allowed, for a total of 47 neutron stars. 
  The principal equilibrium properties are summarized in Table~\ref{tab:equil}.
  Fig.~\ref{fig:equil} shows the mass-radius diagram for all the models 
  computed. The star radius spans a range from $R\sim5$ (in the case of 
  EOS B and G) to $R\sim9-10$ (for EOS L and N). The order of 
  stiffness of the EOS can be estimated, on average,
  as: G$<$B$<$A$<$FPS$<$WFF$<$SLy$<$C$<$O$<$N$<$L. For all the models 
  described by a particular EOS, the compactness $M/R$ increases from 
  about $0.1$ to $0.3$, corresponding to 
  the increase of the star mass and  the decrease of the radius.
  The (total) energy density profile as function 
  of the radial coordinate in the bottom panel of Fig.~\ref{fig:equil} explains,
  as discussed above, that most of the mass is due to matter with density 
  comparable to the central (maximum) density of the star.
  
  \begin{figure*}[t]
    \begin{center}
      \includegraphics[width=0.45\textwidth]{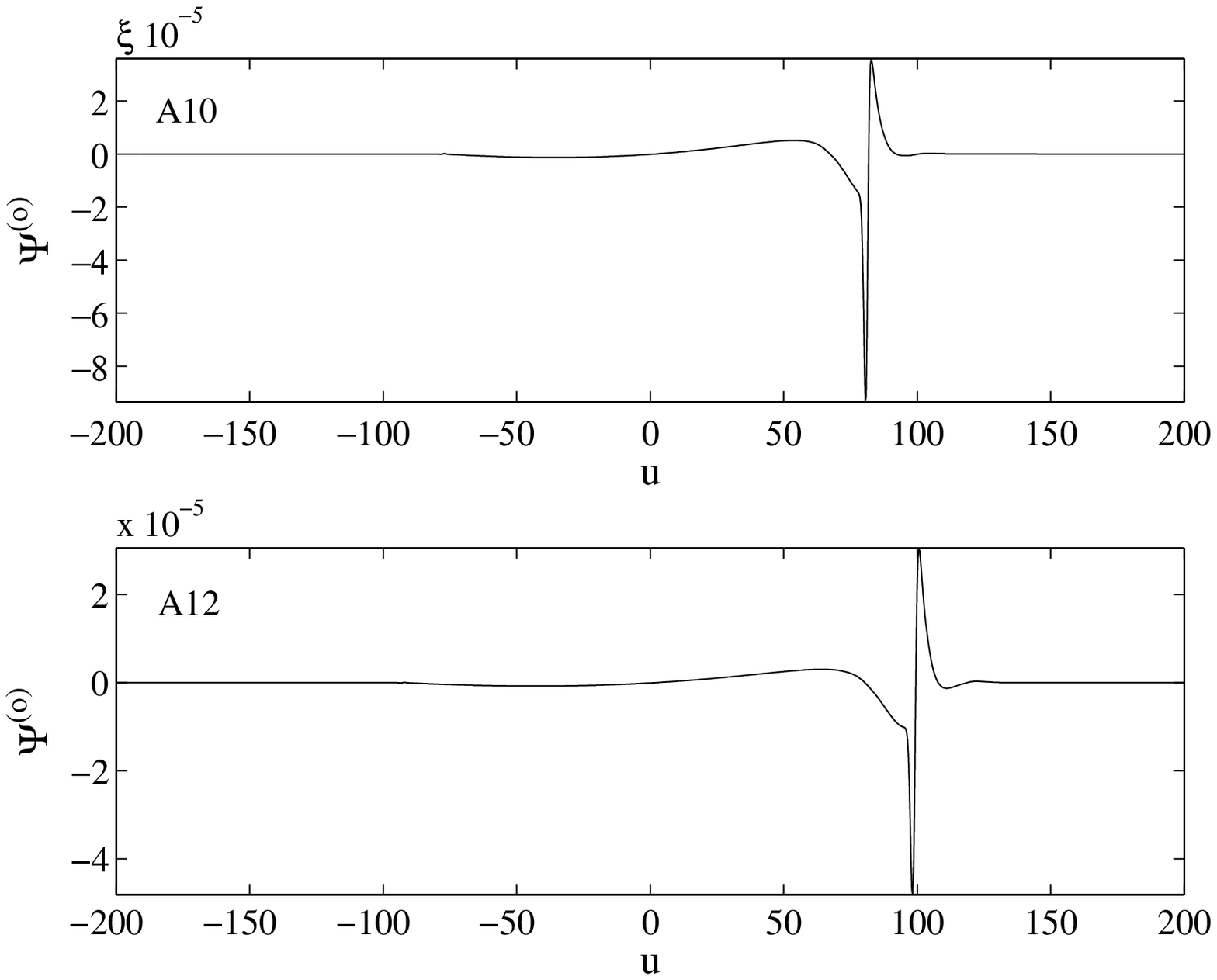}
      \includegraphics[width=0.45\textwidth]{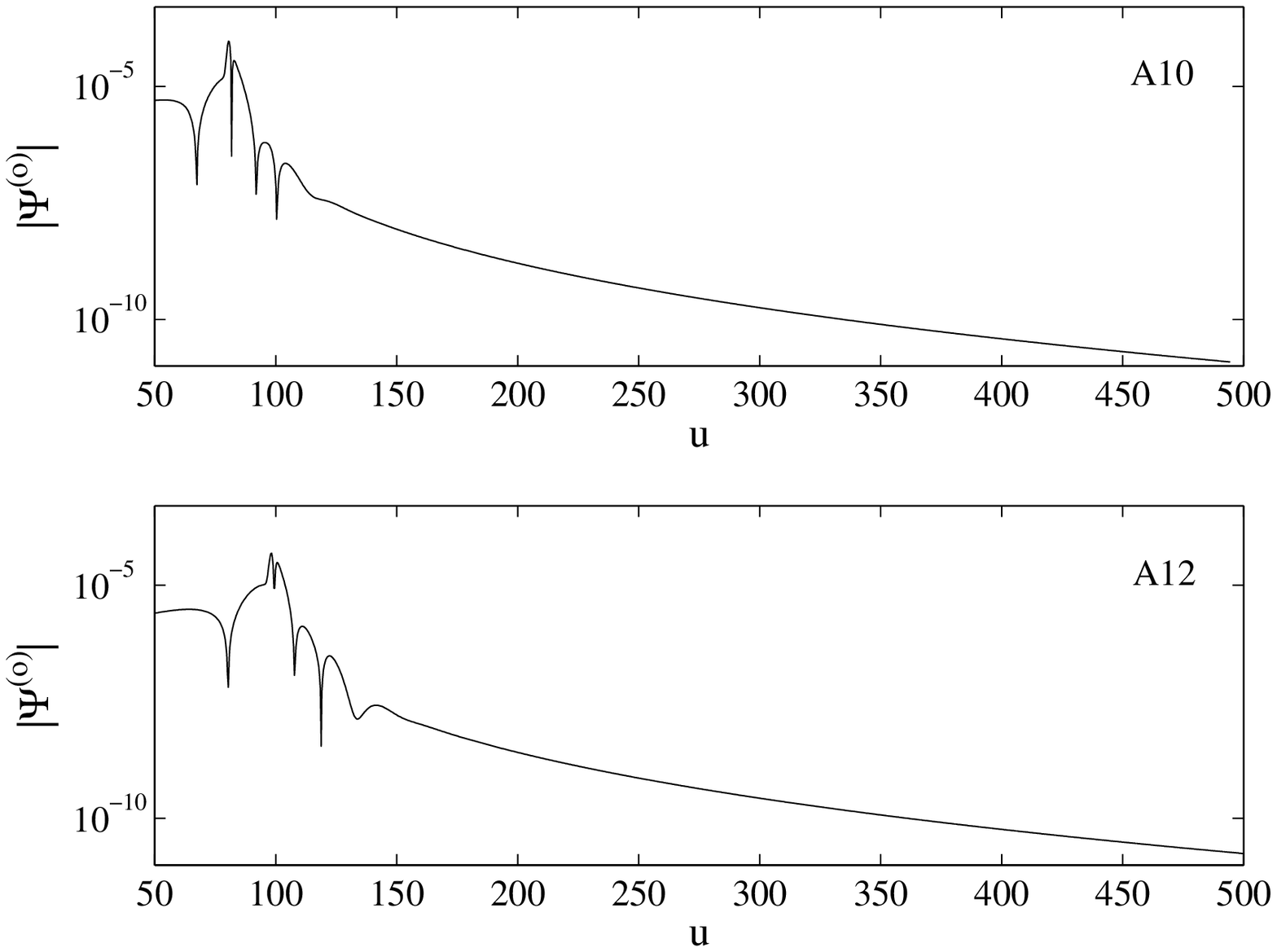}\\
      \includegraphics[width=0.45\textwidth]{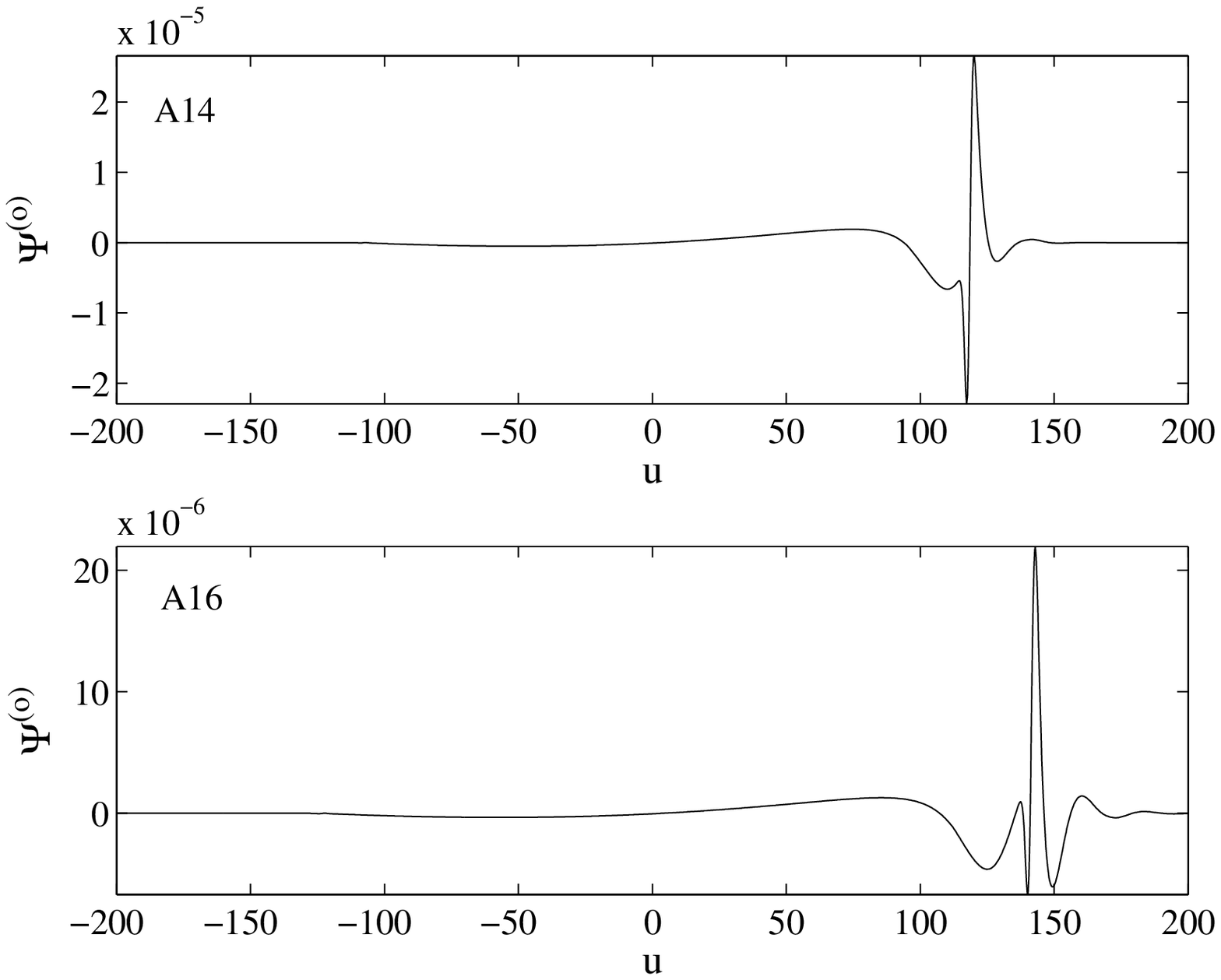}
      \includegraphics[width=0.45\textwidth]{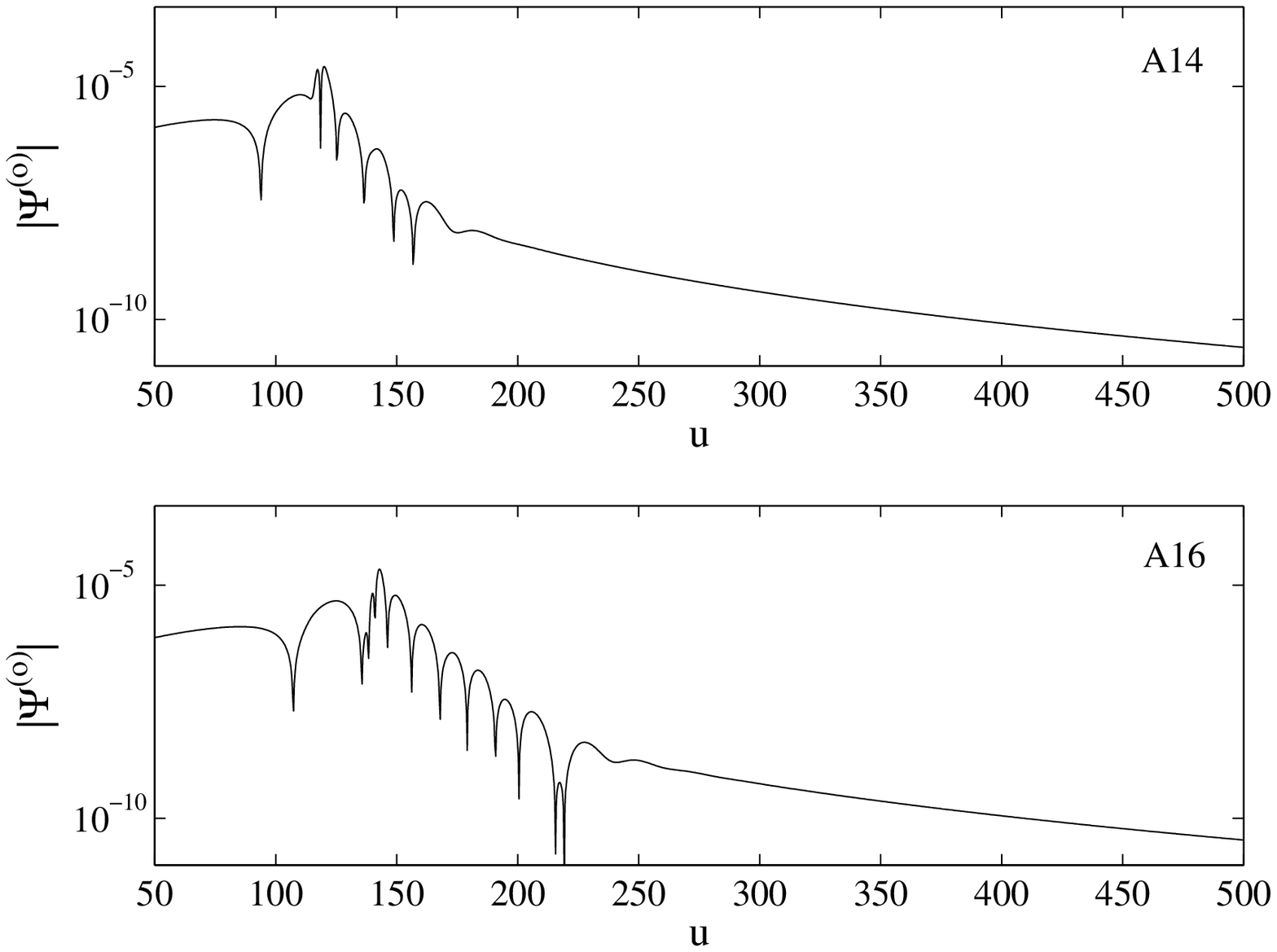}\\
      \includegraphics[width=0.45\textwidth]{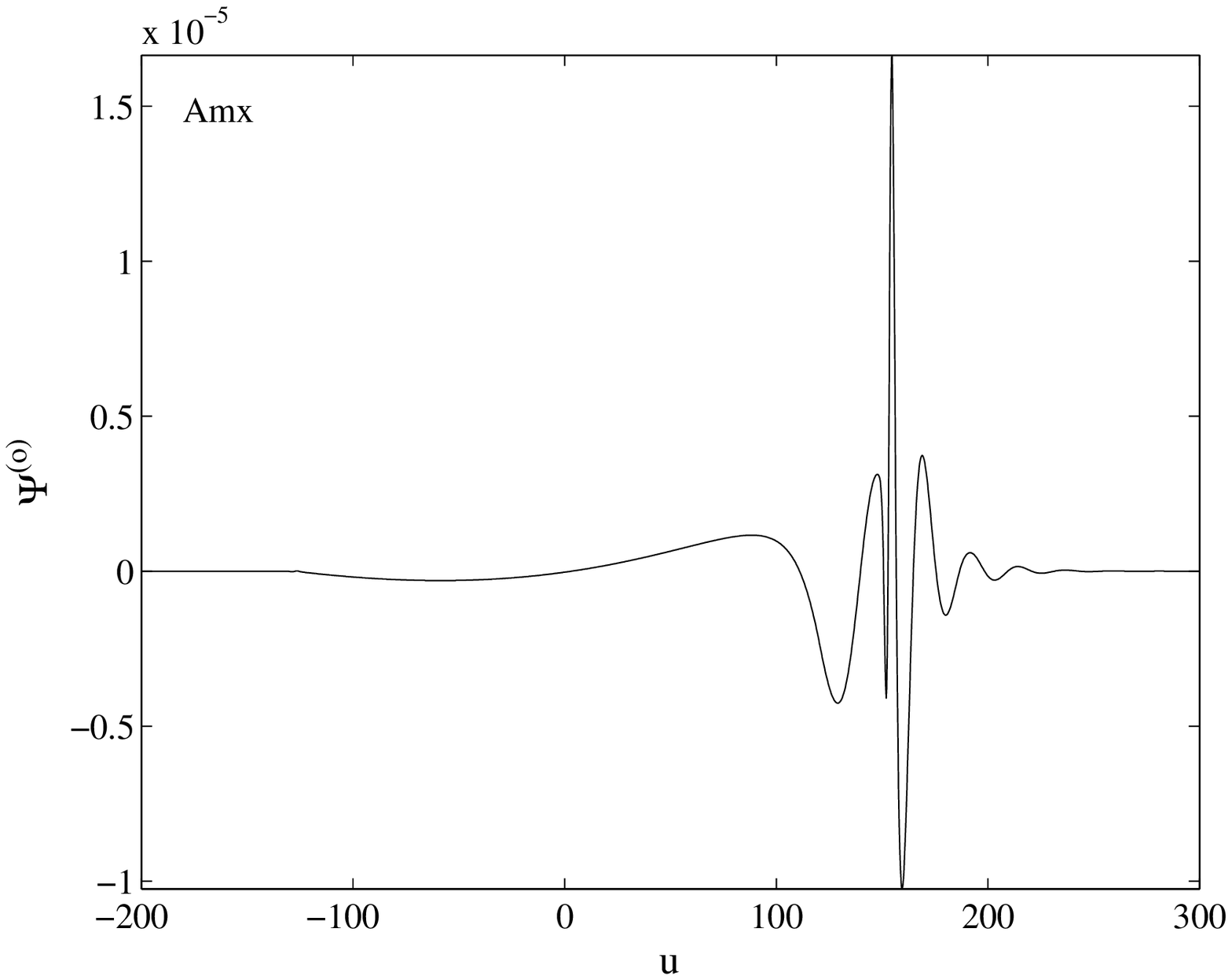}
      \includegraphics[width=0.45\textwidth]{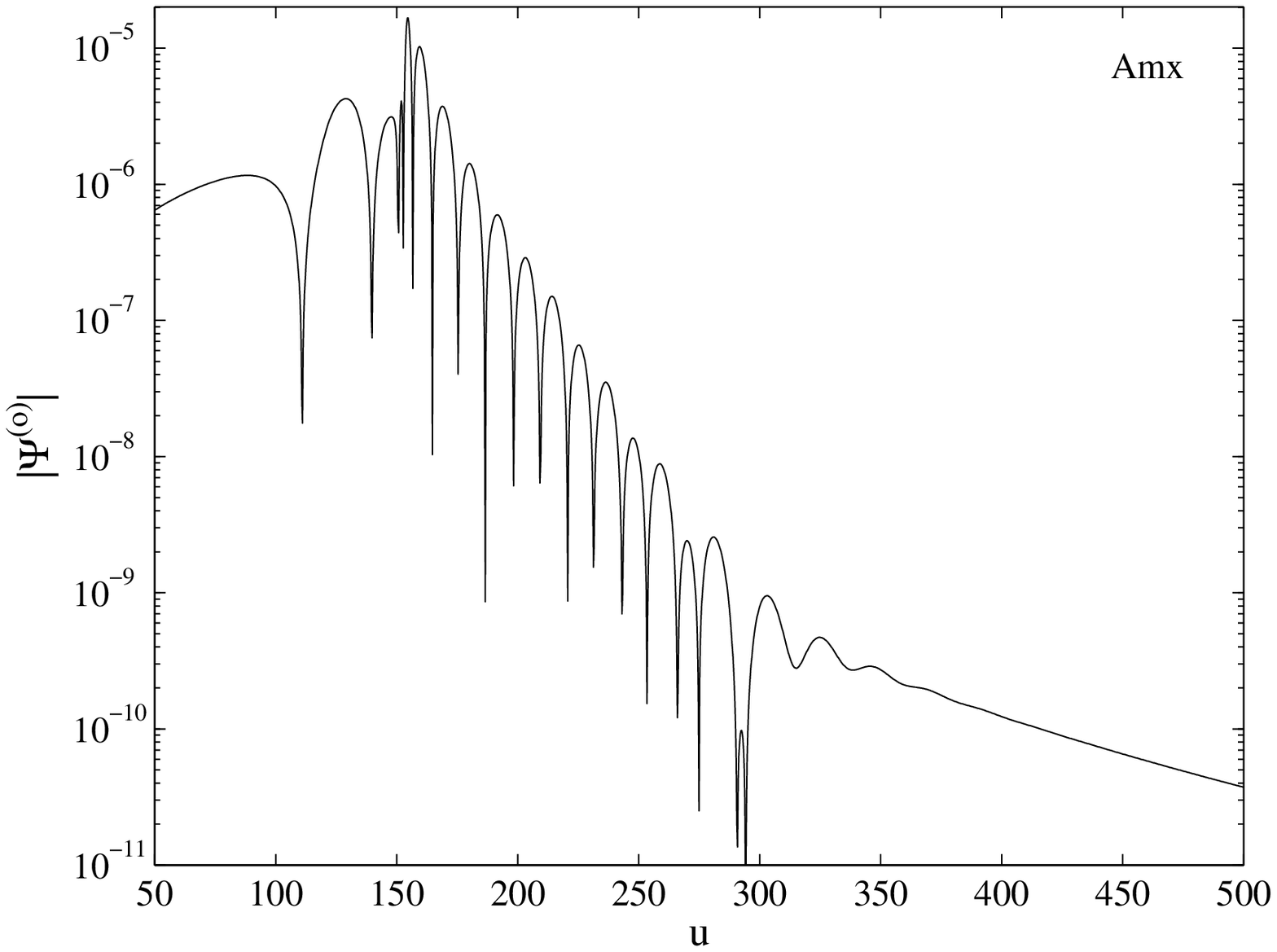}
      \caption{\label{fig:AxialWvfs:eosA} Excitation of $w$-modes in the odd parity waveforms, 
	for different star models with EOS A, generated by the scattering of 
	a Gaussian pulse of GWs with $b=M$. The mass and compactness 
	of the star increases from top to bottom. The presence of $w$-modes is
	more evident for the more compact models. This qualitative behavior
	is common to all EOS.}
    \end{center}
  \end{figure*}

  For each model, we numerically evolve the equations for the odd and 
  even-parity  perturbations described in Sec.~\ref{sec:frmls} using 
  the initial data presented in Sec.~\ref{sec:id}. All the details of 
  the numerical schemes employed can be found in Appendix~\ref{app:num}. 
  In the following sections we will discuss the results obtained
  focusing on the  $\ell=2$ (quadrupole) multipole, as this is the principal 
  responsible of the gravitational wave emission. The waves are extracted at 
  different radii, $r^{\rm obs}=[50,\,100,\,200,\,300]M$. 
  We checked the convergence of the waves and the differencies between the 
  extraction at $r^{\rm obs}=200M$ and $r^{\rm obs}=300M$ are very small, 
  so that we can infer to be sufficiently far away from the source.
  The gravitational waveforms we discuss in the following have 
  always been extracted at the farthest observer, $r^{\rm obs}=300M$ 
  and they are plotted versus observer's retarded time $u=t-r^{\rm obs}_*$.
  
  \subsection{Axial Waveforms}
  \label{sbsec:axial}  
  
  The gravitational waveform that results from scattering of 
  Gaussian pulses of GWs off the odd-parity potential exhibits 
  the well known structure (\emph{precursor-burst-ringdown-tail}~\cite{Davis:1972ud}) 
  analogue to the black holes case (see for example
  Ref.~\cite{Bernuzzi:2008rq} for the case of polytropic EOS). 
  The characteristic signature of the star in the waveform is 
  contained in the ringdown part, which is shaped by high 
  frequencies, (quickly) exponentially damped oscillations: 
  the $w$-modes~\cite{Kokkotas:1992ka}. These modes are pure 
  spacetime vibrations and are the analogue of black hole
  QNMs for relativistic stars~\cite{1999CQGra..16R.159N,lrr-1999-2}.
  
  As a representative case, because the global qualitative features are 
  common to all EOS, Fig.~\ref{fig:AxialWvfs:eosA} exhibits waveforms 
  computed only with EOS A. The compactness of the model increases from 
  top to bottom; the left panels exhibit the waveforms on a linear scale, 
  while the right panels their absolute values on a logarithmic scale.
  Since the damping time increases with the star compactness,
  the maximum mass model, Amx, presents the longest $w$-mode ringdown. 
  On the contrary, model A10 exhibits only a one--cycle, small--amplitude 
  ringdown oscillation that quickly disappears in the power-law tail.

  In principle, an analysis of the frequency content of the axial waveforms by 
  looking at the Fourier spectra or by means of a fit procedure based on a 
  quasi-normal mode template is possible. Let us focus on model Amx, that 
  presents the longest and clearest ringdown waveform. Using the same 
  fit-analysis method of~\cite{Bernuzzi:2008rq} we estimate
  the frequency of the fundamental 
  $w$-modes to be $\nu_w^{(\rm o)}=9452$ Hz, with a damping time 
  $\tau_w^{(\rm o)}\simeq 0.07$ ms. 
  For comparison, we note that a Schwarzschild black hole 
  of the same mass has the fundamental $\ell=2$ frequency and 
  damping time equal to, respectively, $\nu^{\rm BH}=7317$ Hz 
  and $\tau^{\rm BH}=0.09$ ms. We have also computed the energy 
  spectrum of the waveform starting from $u\sim170$  (the first zero after the burst): 
  the spectrum has a single peak centered at a frequency that differs 
  from $\nu_w^{(\rm o)}$ of about a few percents.  
  However, as discussed in~\cite{Bernuzzi:2008rq},
  we found that this information is in general very difficult to 
  extract, especially for the lowest mass models, due to the 
  rapid damping of the modes and their \emph{localization} in a narrow time window. 
  The comparison with frequency domain data (see Ref.~\cite{Bernuzzi:2008rq}
  for polytropic EOS and the discussion in Appendix~\ref{app:num} for
  realistic EOS) shows that the errors on the numbers presented above are 
  of the order of 5\%. The error on the frequencies  
  increases up to about 12\% for models with $M\sim1.4$ and to about 
  20\% for models with  smaller mass. Moreover, the damping times can not 
  be reliably estimated with a fit procedure when they are too short 
  (see Table~\ref{tab:checkfrqs} in Appendix~\ref{app:num}). 
  In summary, although the analysis of the waveforms through a fit procedure 
  works for some particular models, in general it seems uncapable to give 
  numbers as robust and reliable as those provided by a standard 
  frequency domain approach. See for example Ref.~\cite{Benhar:1998au} 
  for details about this approach. 

  \subsection{Polar Waveforms}
  \label{sbsec:polar}  
 
  We start this section presenting together waveforms generated by 
  conformally flat (type 1) and radiative (or non-conformally flat, type 2) 
  initial data. Former studies with polytropic EOS~\cite{Ruoff:2001ux,Nagar:PhD}
  showed that initial data of type 1 determine the excitation of fluid 
  modes only. On the contrary, initial data of type 2 produce a gravitational
  wave signal where both spacetime and fluid modes are present.
  
\begin{figure}[t]
    \begin{center}
      \includegraphics[width=0.50\textwidth]{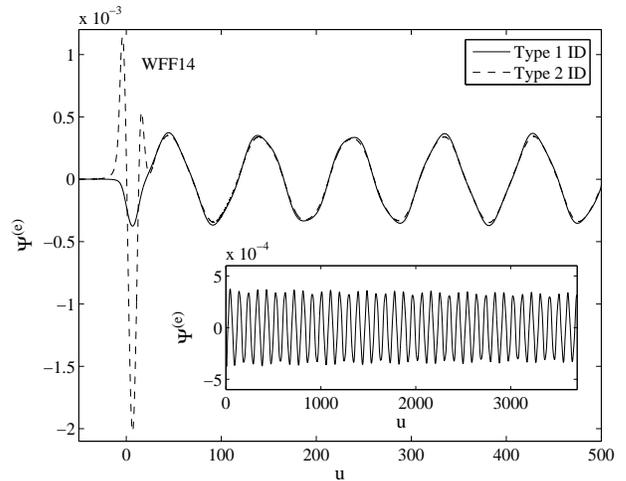}
      \caption{\label{fig:PolarWvfs:WS14:BF} Even parity waveforms for 
	model WFF14 generated by initial data of type 1 (solid line) and
	type 2 (dashed line). In the first case, only fluid modes 
	(the $f$-mode and the first $p$-mode) are present; in the second
	case, $w$-mode oscillations are present at early times 
	($0\lesssim u\lesssim 50$). The inset shows long-term evolution
      (corresponding to a total time of~$\sim20$ ms) used to compute 
	the energy spectrum of Fig.~\ref{fig:PolarWvfs:WS14:HK}.}
    \end{center}
  \end{figure}

\begin{figure}[t]
    \begin{center}
      \includegraphics[width=0.50\textwidth]{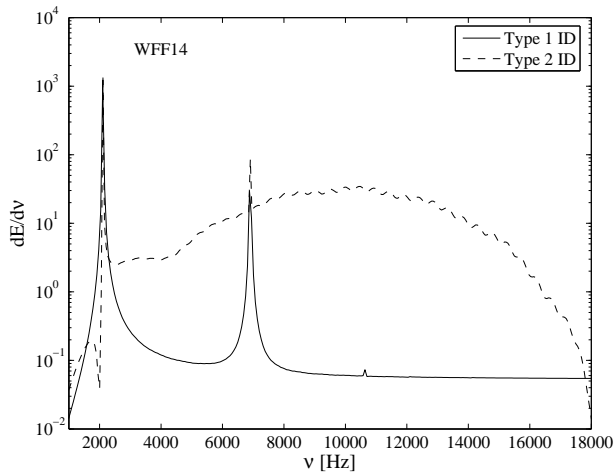}
      \caption{\label{fig:PolarWvfs:WS14:HK} Energy spectra of
	the type 1 (solid line) and type 2 (dashed line) initial data
	evolution of Fig.~\ref{fig:PolarWvfs:WS14:BF}. The two 
	narrow peaks at 2126 Hz and 6909 Hz correspond to the $f$-mode and 
	the first $p$-mode frequencies. The wide peak at $\simeq 11$~kHz
	corresponds to $w$-mode excitation.}
    \end{center}
  \end{figure}

  As expected, this qualitative picture is confirmed also for realistic EOS.
  Figure~\ref{fig:PolarWvfs:WS14:BF} exhibits the waveform $\Psi^{(\rm e)}$ 
  for the representative model WFF14. For conformally flat initial data 
  (solid line) the figure shows that the Zerilli-Moncrief function 
  oscillates at (mainly) one frequency, of the order of the kHz. The 
  corresponding energy spectrum (solid line in
  Fig.~\ref{fig:PolarWvfs:WS14:HK}) reveals that the signal is
  in fact dominated by the frequency $\nu_f=2126$ Hz, but 
  there is also a second peak at $\nu_p=6909$ Hz. This two frequencies 
  are recognized as those of the fundamental fluid mode $f$ and of the 
  first (pressure) $p$-mode.

  For non-conformally flat initial data (dashed line in Fig.~\ref{fig:PolarWvfs:WS14:BF})
  the first part of the signal, i.e. $0\lesssim u\lesssim 50$, is dominated by
  a high-frequency and strongly-damped oscillation typical of curvature modes.
  For $u\gtrsim50$, the type 1 and type 2 waveforms are practically
  superposed. The corresponding energy spectrum (dashed line in Fig.~\ref{fig:PolarWvfs:WS14:HK}
  has, superposed to the two narrow peaks of the fluid modes, a wide 
  peak centered at higher frequency ($\sim 11$ kHz) that is typical 
  of the presence of spacetime excitation~\cite{lrr-1999-2}.~\footnote{Note 
  that, in order to obtain the cleanest fluid-mode peaks, in the Fourier 
  transform of type 1 waveform we discarded the first four GW cycles,
  which are contaminated by a transient due to the initial excitation
  of the system. On the other hand, for type 2 waveform,  we considered 
  the full time-series. In this case, it is not possible to cut the first
  part of the signal because it also contains the curvature mode 
  contribution that we want to analyze.}

\begin{figure}[t]
    \begin{center}
      \includegraphics[width=0.5\textwidth]{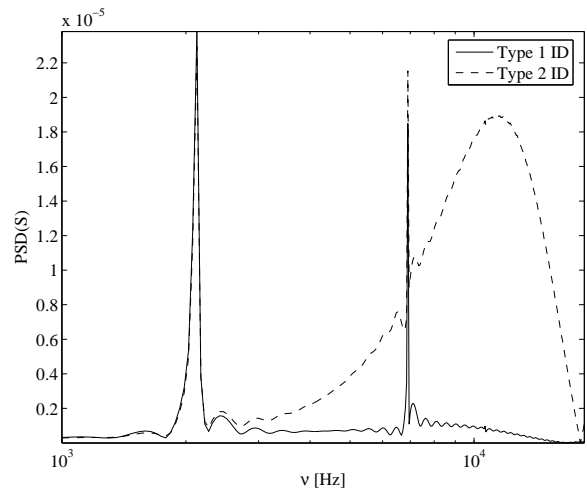}\\      
      \caption{\label{fig:PSD:S} Power spectrum of the variable $S$ for model WFF14
	in the case of initial data of type 1 and 2.}
    \end{center}
  \end{figure}

  The information of Fig.~\ref{fig:PolarWvfs:WS14:HK} is complemented
  by the Fourier spectrum of the metric variable $S$ in Fig.~\ref{fig:PSD:S}.
  In contrast to the fluid variable $H$, which contains only narrow 
  peaks for both kind of initial data, for type 2 waveforms 
  the metric variable $S$ also exhibits a broad peak which is
  absent for type 1 initial data. Note, that the picture that 
  we have discussed so far for model WFF14 remains qualitatively 
  unchanged for all the other EOS. 

\begin{figure}[t]
    \begin{center}
      \includegraphics[width=0.5\textwidth]{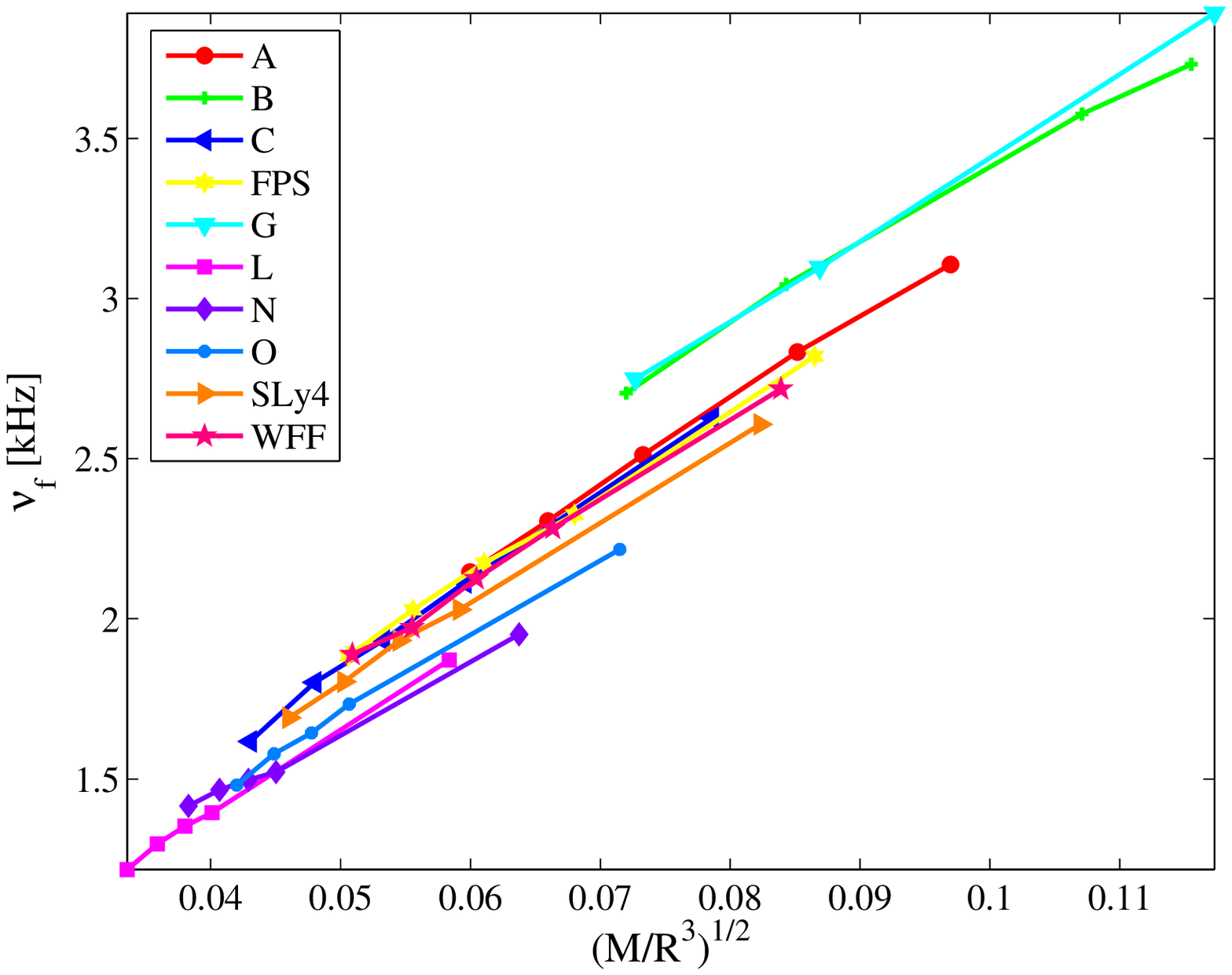}
      \includegraphics[width=0.5\textwidth]{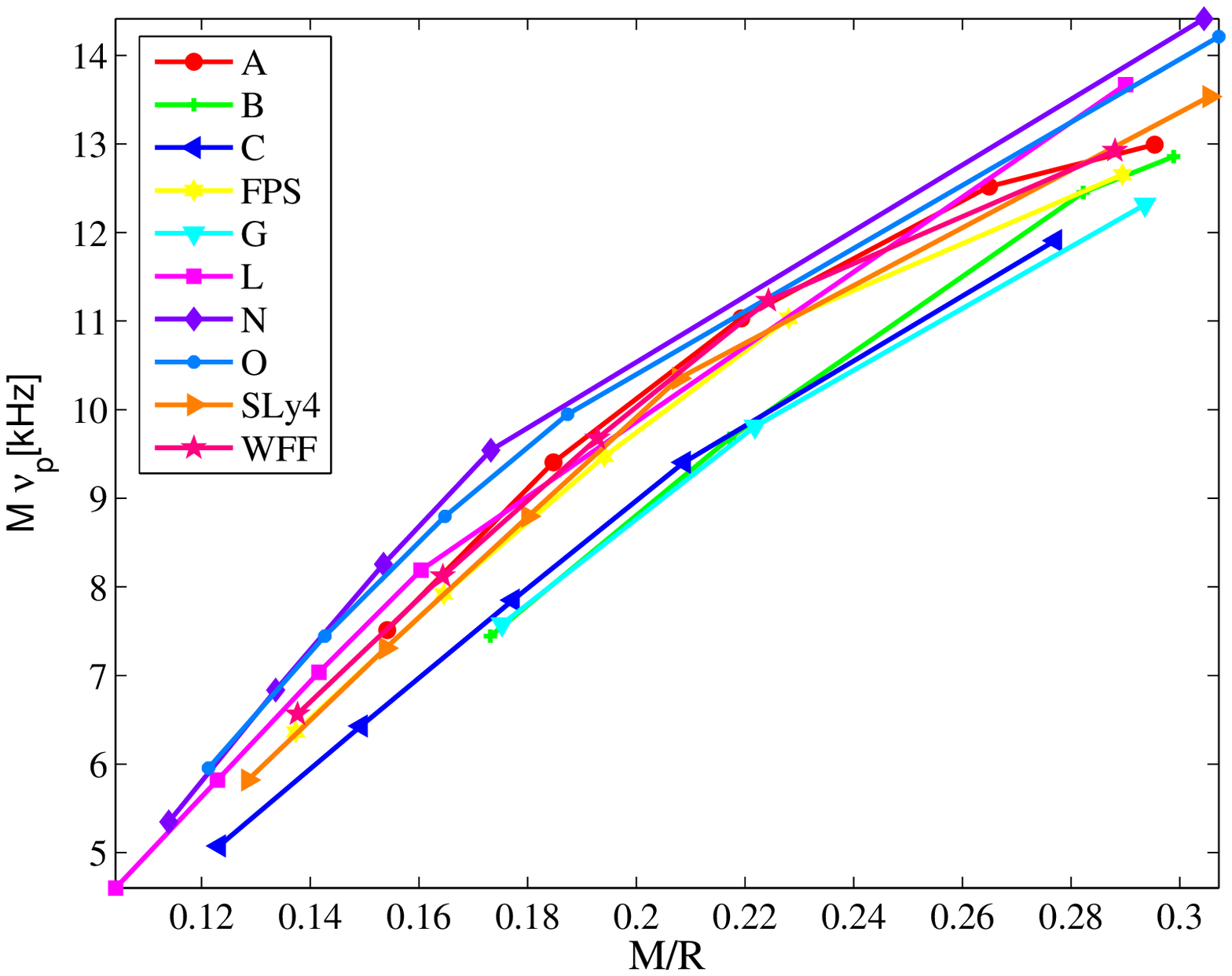}
      \caption{\label{fig:frqs} Comparison between the oscillation
	properties of models computed with different EOS. Top panel: 
	$f$-mode frequencies as a function of $\sqrt{M/R^3}$.
	Bottom panel: the first $p$-mode frequency (multiplied by the
	mass $M$) as a function of the compactness $M/R$.}
    \end{center}
  \end{figure}

  Since our numerical scheme allows us to evolve the system in time
  as long as we wish, we can produce very long time-series to accurately 
  extract, via Fourier analysis, the fluid mode frequencies. 
  We did this analysis systematically for all the models considered.
  In Tables~\ref{tab:frqsA}-\ref{tab:frqsWFF} in Appendix~\ref{app:num} 
  we list the frequencies of the $f$-mode and the first $p$-mode 
  extracted from the energy spectra. By comparison with the published
  frequencies of Andersson and Kokkotas~\cite{Andersson:1997rn} for
  some models with EOS A (obtained via frequency domain calculations), 
  we estimate that the errors on our values are tipically smaller than 1\%.
  For a fixed EOS, the frequencies increase with the star compactness.
  For a model of given mass, the $f$-mode frequency generally decreases 
  if the EOS stiffens. The same (on average) is true for the first $p$-mode
  frequency. Following Ref.~\cite{Andersson:1997rn}, we present in
  Fig.~\ref{fig:frqs} the frequencies that we have computed 
  as a function of the mean density $\sqrt{M/R^3}$ ($f$-mode) 
  and of the compactness $M/R$ ($p$-mode) of the star.
  Globally, they show a very good quantitative agreement 
  with previously published results calculated by means of a 
  standard frequency domain approach~\cite{Andersson:1997rn}.   

\begin{figure}[t]
    \begin{center}
      \includegraphics[width=0.50\textwidth]{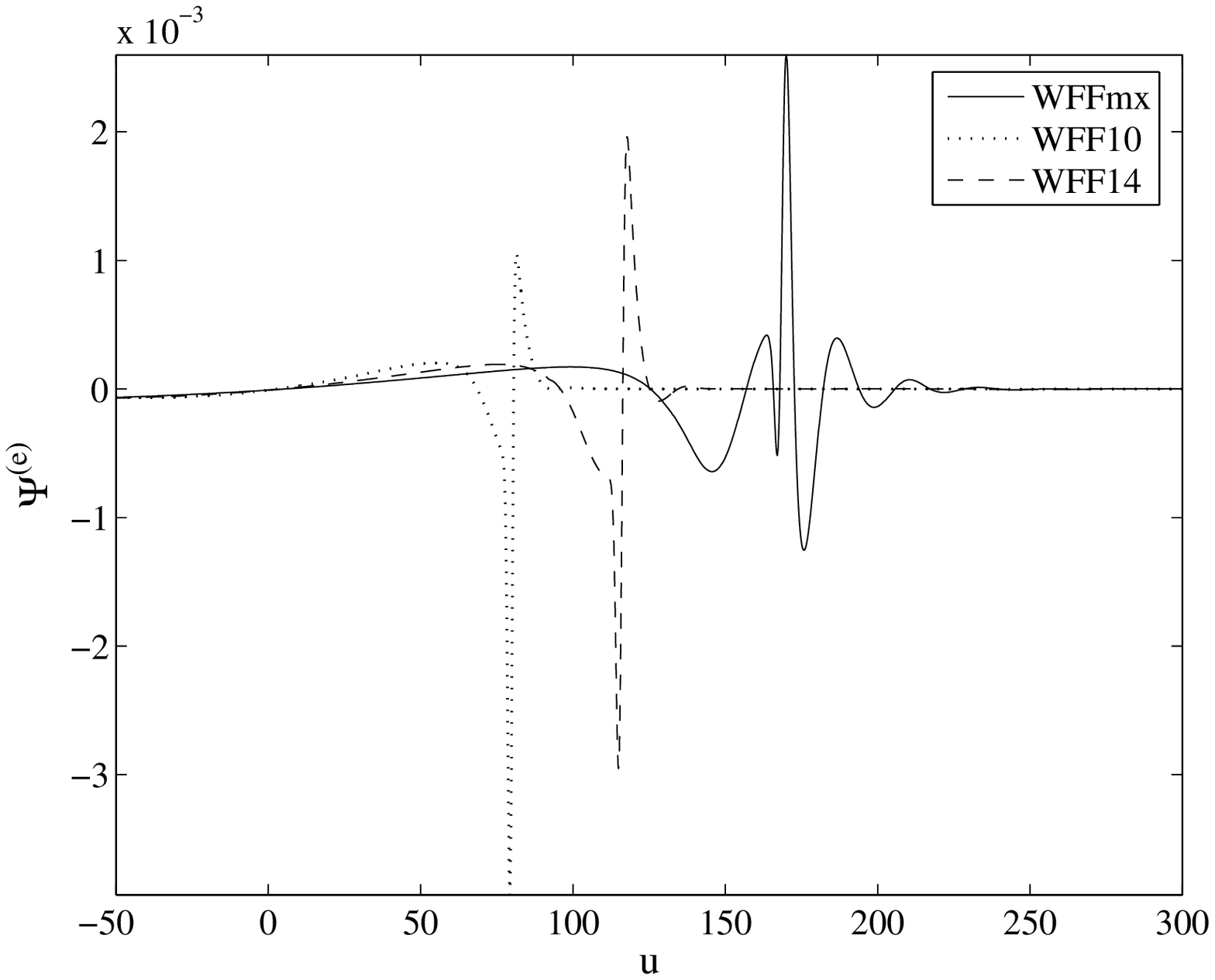}\\
      \includegraphics[width=0.50\textwidth]{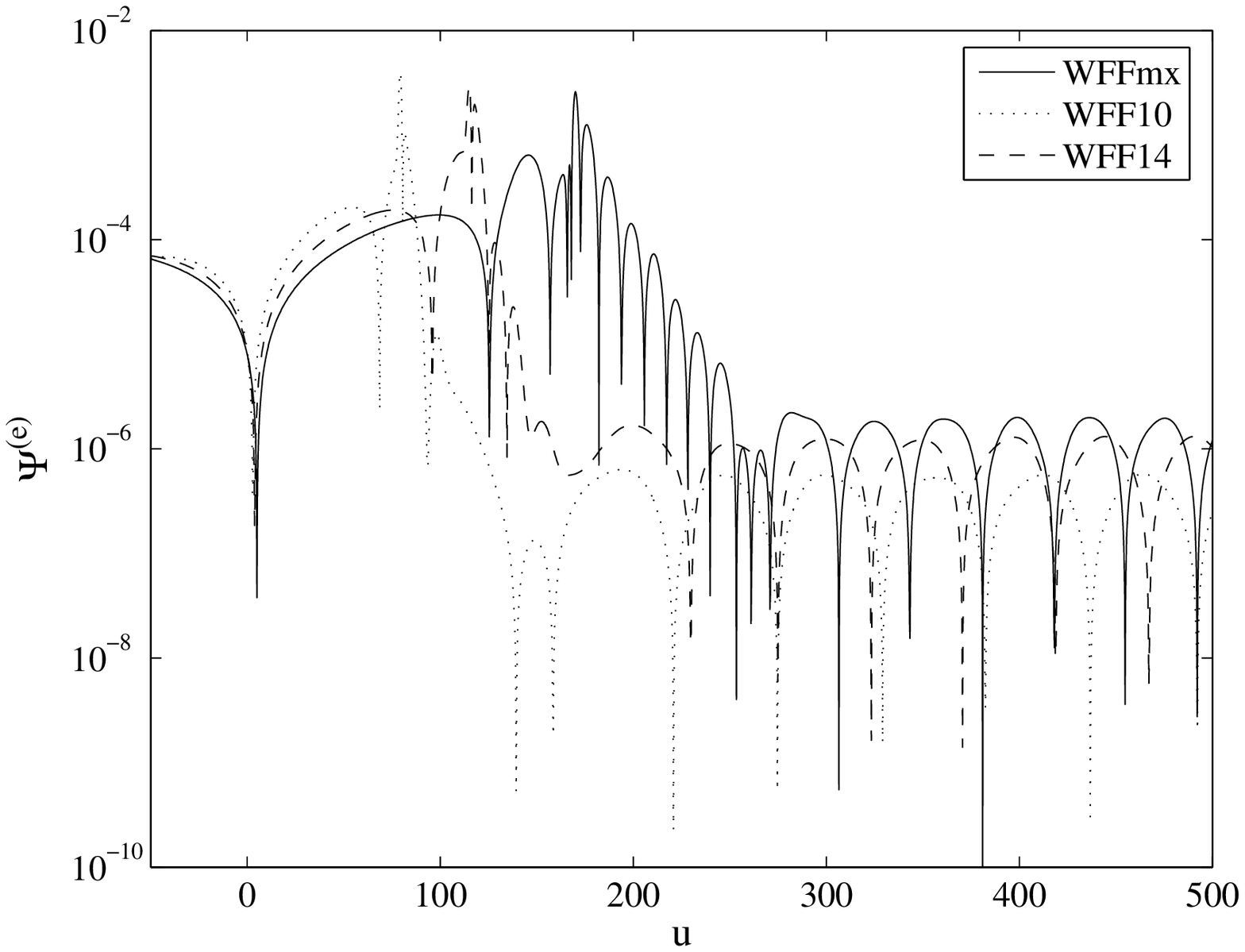}
      \caption{\label{fig:PolarWvfs:WS:ZG} Excitation of even-parity $w$-modes 
	for some WFF models from given initial data of type 3. As in the
	odd-parity case (compare with Fig.~\ref{fig:AxialWvfs:eosA}), the ringdown is
        more pronounced for more compact models.}
    \end{center}
  \end{figure}

  We conclude this section by discussing the waveforms generated by initial
  data of type 3. This kind of ``scattering-type'' initial condition 
  constitutes the even-parity analogue of that discussed in 
  Sec.~\ref{sbsec:axial} for the odd-parity case.  
  In Fig.~\ref{fig:PolarWvfs:WS:ZG} we show the $\Psi^{(\rm e)}$ 
  waveforms from 3 models of EOS WFF. The waveforms in the top panel 
  of the figure are very similar to those of the left panels of 
  Fig.~\ref{fig:AxialWvfs:eosA}. The logarithmic scale (bottom panel)
  highlights the main qualitative difference: i.e., fluid-mode oscillations 
  are present, in place of the nonoscillatory tail, after the $w$-mode 
  ringdown. Note that, in principle, the tail will emerge in the signal 
  after that all the fluid modes have damped (i.e., on a time scale of a few seconds).

  The process of $w$-mode excitation is instead exactly the same 
  as for the odd parity case: the ringdown phase is longer 
  (and thus clearly visible) for the more compact models.
  The frequencies are also very similar.
  For example, for model WFFmx (the one discussed in the figure) 
  we have $\nu_w^{(\rm e)}=8638$ Hz and a damping time $\tau_w^{(\rm e)}
  \simeq0.05$ ms, while for model 
  Amx $\nu_w^{(\rm e)}=9798$ Hz and $\tau_w^{(\rm e)}\simeq0.05$ ms 
  (to be compared with $\nu_w^{(\rm o)}=9452$ and $\tau_w^{(\rm o)}\simeq
  0.07$ ms).

  \section{Conclusions}
  \label{sec:conc}  
  
  In this work we have discussed the time-evolution of nonspherical
  (matter and gravitational) perturbations of nonrotating neutron
  stars described by a large sample of realistic EOS.
  The current study extends the work of Allen et al.~\cite{Allen:1997xj} 
  and Ruoff~\cite{Ruoff:2001ux}, who focused essentially on polytropic,
  but one, EOS models. We have used an improved version of a 
  recently developed 1D perturbative code~\cite{Nagar:2004ns} that has 
  been thoroughly tested and used in the literature~\cite{Nagar:PhD,Nagar:2004av,Passamonti:2007tm}.
 
  The main, new result, presented here is that our {\it constrained} 
  numerical scheme allows us to stably evolve the even-parity perturbation equations
  without introducing any ``special'' coordinate change, as it was necessary 
  in Ref.~\cite{Ruoff:2001ux}. 
  In addition, despite the EOS that we consider are very different, 
  the outcome of our computations is fully consistent (as expected)
  with previous studies involving polytropic EOS. In particular:
  (i) for even-parity perturbations, if the initial configuration involves 
  a fluid excitation, (type~1 and~2 initial data), the Zerilli-Moncrief function 
  presents oscillations at about 2-3 kHz due to the excitation of the fluid QNMs of the star; 
  (ii) if we set $S\neq 0$ at $t=0$ (type 2, i.e. the non-conformally 
  flat condition is imposed), high frequencies, strongly damped $w$-mode 
  oscillations are always present in the waveforms; (iii) the $w$-mode 
  excitation is generally weak, but it is less weak the more compact
  the star model is; consistently, (iv) for scattering-like initial data 
  in both the odd and even-parity case the presence of $w$-modes 
  is more striking the higher is the compactness of the star \footnote{Note 
  that type 3 initial data are non conformally flat as well, 
  since $\chi\neq 0$.}; even-parity fluid modes are typically weakly excited in this case.

  Thanks to the long-term and accurate evolutions we can 
  perform, we extracted the fluid mode frequencies from the 
  Fourier transform of time series of the waves with an 
  accuracy comparable to that of frequency domain codes. 
  For what concern the frequencies and damping times of spacetime modes, 
  pretty good estimates can be obtained 
  when damping times are not too short; i.e., for the more compact models.
  When these frequencies will be revealed in a gravitational 
  wave signal, they will hopefully provide useful information 
  on the internal structure of neutron stars. 
  In particular, recognizing both fluid and $w$-modes 
  in the signal could permit, in principle, to estimate the 
  values of mass and radius of the NS and thus 
  to put strong constraints on the EOS model~\cite{Andersson:1997rn,Benhar:2004xg,Ferrari:2007dd}.
  We have limited our analysis to the first two fluid modes 
  because they are the most responsible for gravitational waves emission.
  We checked that, by changing the initial fluid perturbation,
  our evolutionary description permits also to easily capture the
  frequencies of higher overtones~\cite{Nagar:2004av}.

  In addition, despite all the approximation that we have introduced 
  (initial data, no rotation, no magnetic fields),
  we believe that the approach to NS oscillations described in 
  this paper can provide physical information complementary to that available
  from GR nonlinear evolutions. In particular, it can also be used 
  to provide useful test-beds for GR nonlinear codes and must be 
  seen as a first step to compare/constrast with nonlinear simulations
  of neutron star oscillations.
  In conclusion, since we can freely specify the initial data of the metric 
  and matter variables at initial time, one can also think to use 
  our present tool to evolve further in time an almost spherical 
  configuration that is the outcome of a long-term numerical 
  relativity simulation. A perturbative evolution like the one we 
  discussed here (possibly complemented by a complex 3D (magneto)-hydrodynamics source, 
  as an improvement of the approach discussed in Ref.~\cite{Nagar:2004ns})  
  could then start when the 3D fully nonlinear simulation ends.  
  
  
  \section*{Acknowledgments}

  We are grateful to T.~Damour, V.~Ferrari, P.~Haensel, B.~Haskell, 
  K.~Kokkotas, A.~Potekhin and N.~Stergioulas for critical readings of the manuscript. 
  We thank R.~De Pietri for discussions and assistance during the 
  development of this work. The EOS tables were taken from~\cite{Haensel:2004nu,NSG,RNS}. 
  All computations performed on the {\tt Albert} Beowulf clusters at the 
  University of Parma. The activity of AN at IHES is supported by INFN. 
  SB gratefully acknowledges support of IHES, where part of this work 
  was done. The commercial software Matlab has been used in the 
  preparation of this work.

 
  \appendix   
  
  \section{Table Interpolation}
  \label{app:int}  
    
  Here we describe the method used to interpolate the tables of the EOS.
  The interpolation scheme is based on Hermite polynomials and was
  introduced in Ref.~\cite{Swesty:1996}. We follow~\cite{Nozawa:1998ak} 
  by using 3rd order (cubic) polynomials and we list here the relevant 
  formulas for completeness. Consider a function $y(x)$, given the 
  tables of $y_j$ and $y'_j$ and a point $x_{\rm j}<x<x_{\rm j+1}$ 
  the interpolated value $y(x)$ is:
  \begin{align}
    y(x) & = y_i H_0(w)+ y_{i+1}H_0(1-w) \\
         & + \left(\frac{dy}{dx}\right)_j\Delta x_j H_1(w)\nonumber\\
         & + \left(\frac{dy}{dx}\right)_{j+1}\Delta x_j H_1(1-w),\nonumber
  \end{align}
  where 
  \begin{align}
    \Delta x_j &\equiv (x_{j+1}-x_j), \\
          w(x) &\equiv \frac{x-x_j}{\Delta x_j}
  \end{align}
  and
  \begin{align}
    H_0(w) &= 2w^3 -3w^2 +1,\\
    H_1(w) &= w^3 -2w^2 +w
  \end{align}
  are the cubic Hermite functions. The principal properties of this method are 
  that for $x\rightarrow x_j$
  \begin{align}
    y(x) &\rightarrow y_j,\\
    y'(x) &\rightarrow y'_j .
  \end{align}
  
  The EOS are usually given as 3 column tables with the 
  values for $n$, $\e$ and $p$.
  In this work we need to compute $\e(p)$ and the speed of sound $\cs$. 
  The thermodynamics consistency can be achived by computing, 
  for a given value of $p$, first $n(p)$ and then $\e(n)$ 
  imposing the derivative through Eq.~\eqref{eq:1lawther}. Finally 
  the speed of sound is computed consistently with the interpolation. 
  To obtain more accurate numerical data, we perform such calculations
  not directly on the functions, but taking logarithms.
  Violation of the first law of thermodynamics are typically less 
  than 0.1\%. We tested also other interpolation schemes, i.e. linear 
  and spline interpolation, that in general gives violation of the 
  thermodynamic principle of some percents.
  
  As a check of the implementation, we evolved a polytropic model 
  both with the analytic EOS and with tables of different numbers 
  of entries. The results perfectly agree and we did not find any 
  dependence on the tabulated points. In addition, the use of other 
  interpolation schemes did not produce significant differences: 
  this fact suggests that the global numerical errors of the code 
  are dominant over the errors related to the violation of the 
  thermodynamics principle. In the case of some tables 
  (EOS FPS, SLy4 and L), we found that ``high-order'' 
  interpolation (cubic Hermite and spline) did not permit an accurate 
  reconstruction of the sound speed. This was due to spurious oscillations 
  introduced by the high order derivatives. As a consequence, the 
  code gave unphysical results. In all these cases, we adopted 
  the linear interpolation.

  
  \section{Numerical Details, code tests and mode frequencies}
  \label{app:num}  
  
  The code we employ in this work is a development of that described 
  in~\cite{Nagar:PhD} and successfully used in many 
  works~\cite{Nagar:2004av,Nagar:2004ns,Passamonti:2005cz,Passamonti:2007tm}.
  
  The TOV equations~\eqref{eq:TOV} are integrated numerically 
  (from the center outward),
  for a given central pressure $p_{\rm c}$ (see Table~\ref{tab:equil}),
  using a standard fourth-order Runge-Kutta integration scheme with
  adaptive step size.
    
  To evolve numerically the perturbations equations, we introduce
  an evenly spaced grid in $r$ with uniform spacing $\Delta r$ and 
  we adopt finite differencing approximation schemes for the derivatives. 
  In particular, in the construction of the the computational grid the origin
  $r=0$ is excluded and the first point is located at $r=\Delta r/2$. 
  The resolution is measured as the number of point $J_i$ inside the star 
  radius. The star surface is located at a cell center 
  $R=\Delta r/2 + (J_i-1)\Delta r$.

  The hyperbolic evolution equations for $S$ and $H$ are all 
  solved with standard, second order convergent in time and 
  space, leapfrog algorithm. For the evolution of the even-parity equations, 
  the Hamiltonian constraint is used to update, at every time step,
  the variable $k$. This elliptic equation is discretized in space at 
  second order and reduced to tridiagonal linear system that is inverted.
  For this reason the evolution scheme of the polar perturbations 
  can be considered a \emph{constrained evolution}. The inner boundary 
  conditions of Eqs.~\eqref{eq:BC1}-\eqref{eq:BC3} are implemented by 
  setting to zero the variables at the first grid point. At the outer
  boundary, standard radiative Sommerfeld conditions are imposed.
  
  The Zerilli-Moncrief function has been obtained in two (independent) ways.
  On the one hand, it has been computed from $S$ ($\chi$) and $k$ using
  Eq.~\eqref{eq:ZM} for every value of $r$. On the other hand, it has been
  computed using Eq.~\eqref{eq:ZM} only at the star surface ($r=R$, the
  matching point) and then evolved using Eq.~\eqref{eq:ZMequation}. 
  The second method has been used only as an independent consistency check
  and all the results discussed in this paper are obtained using the first
  method.

  \begin{figure}[t]
    \begin{center}
      \includegraphics[width=0.5\textwidth]{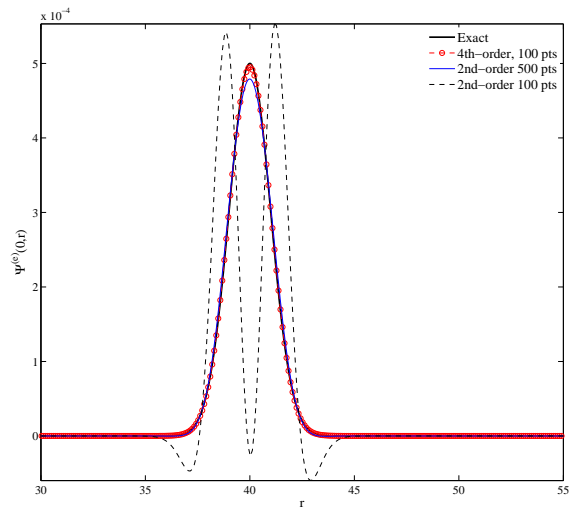}
      \caption{\label{fig:zerilli}Convergence of the Zerilli function
      for initial data of type 3 for a polytropic EOS model. 
      See text for discussion.}
    \end{center}
  \end{figure}

  \begin{table}[t]
    \caption{\label{tab:checkfrqs} Frequencies $\nu$ and damping times $\tau$ 
      of fluid and spacetime modes of some models of Andersson and 
      Kokkotas~\cite{Andersson:1997rn}, described by EOS~A, computed 
      via our approach. The frequencies are expressed in Hz and the damping 
      times in ms. For the sake of comparison, we report also the values of 
      their Table~A.1 with the ``AK'' superscript.}
    \begin{ruledtabular}
      \begin{tabular}{lcccccccc} 
	$M$ & $\nu_f$ & $\nu_f^{\rm AK}$ & $\nu_p$ & $\nu_p^{\rm AK}$ & $\nu_w$ &
	$\nu_w^{\rm AK}$ & $\tau_w$ & $\tau_w^{\rm AK}$\\
	\hline
	1.653 & 3080 & 3090 & 7825 & 7838 & 9342  & 9824 & 0.062 & 0.064\\
	1.447 & 2580 & 2579 & 7843 & 7818 & 10038 & 11444 & 0.057 & 0.027\\
	1.050 & 2183 & 2203 & 7555 & 7543 & 11267 & 14328 & 0.059 & 0.017\\
      \end{tabular}
    \end{ruledtabular}
  \end{table}

 \begin{table}[t]
    \caption{\label{tab:frqsA} Frequencies of the fluid modes 
     for EOS A, B, C, FPS and G. From left to right the 
     columns report: the name of the model, the EOS type, 
     the $f$-mode frequency and the first $p$-mode frequency.}
    \begin{ruledtabular}
      \begin{tabular}{cccc}
       Model & EOS & $\nu_f$ [Hz] & $\nu_p$ [Hz] \\
	\hline
       A10 & A & 2146 & 7444\\
       A12 & A & 2305 & 7816\\
       A14 & A & 2512 & 7873\\
       A16 & A & 2833 & 7823\\
       Amx & A & 3107 & 7848\\
       \hline
       B10 & B & 2705 & 7440\\
       B12 & B & 3044 & 8061\\
       B14 & B & 3577 & 8893\\
       Bmx & B & 3732 & 9091\\
       \hline
       C10 & C & 1617 & 5052\\
       C12 & C & 1802 & 5349\\
       C14 & C & 1933 & 5606\\
       C16 & C & 2114 & 5876\\
       Cmx & C & 2627 & 6421\\
       \hline
       FPS10 & FPS & 1884 & 6326\\
       FPS12 & FPS & 2028 & 6590\\
       FPS14 & FPS & 2174 & 6764\\
       FPS16 & FPS & 2325 & 6891\\
       FPSmx & FPS & 2820 & 7031\\
       \hline
       G10 & G & 2750 & 7510\\
       G12 & G & 3097 & 8164\\
       Gmx & G & 3891 & 9080
    \end{tabular}
\end{ruledtabular}
    \end{table}

 \begin{table}[t]
    \caption{\label{tab:frqsWFF} Frequencies of the fluid modes for 
     EOS L, N, O, SLy and WFF. From left to right the columns report:
    the name of the model, the EOS type, the $f$-mode frequency and the
    first $p$-mode frequency.}
    \begin{ruledtabular}
     \begin{tabular}{cccc}
       Model & EOS & $\nu_f$ [Hz] & $\nu_p$ [Hz] \\
       \hline
       L10 & L & 1217 & 4599\\
       L12 & L & 1297 & 4850\\
       L14 & L & 1353 & 5025\\
       L16 & L & 1395 & 5115\\
       Lmx & L & 1871 & 5109\\
        \hline
       N10 & N & 1415 & 5324\\
       N12 & N & 1466 & 5694\\
       N14 & N & 1497 & 5893\\
       N16 & N & 1522 & 5960\\
       Nmx & N & 1952 & 5470\\
        \hline
       O10 & N & 1481 & 5924\\
       O12 & N & 1578 & 6200\\
       O14 & N & 1643 & 6283\\
       O16 & N & 1734 & 6217\\
       Omx & N & 2217 & 5969\\
	\hline
       SLy10 & SLy & 1691 & 5818\\
       SLy12 & SLy & 1804 & 6087\\
       SLy14 & SLy & 1932 & 6279\\
       SLy16 & SLy & 2029 & 6468\\
       SLymx & SLy & 2607 & 6601\\
	\hline
        WFF & WFF10 & 1889 & 6544\\
        WFF & WFF12 & 1973 & 6766\\
        WFF & WFF14 & 2126 & 6909\\
        WFF & WFF16 & 2282 & 7016\\
        WFF & WFFmx & 2718 & 7016
     \end{tabular}
   \end{ruledtabular}
 \end{table}

  In some situations, we found the convergence of the Zerilli-Moncrief 
  computed from Eq.~\eqref{eq:ZM} particularly delicate. For example, in case of
  type 3 initial data (an even-parity Gaussian pulse of gravitational
  radiation) we realized that a very accurate computation of the radial
  derivative $k_{,r}$ is needed to obtain an accurate and reliable
  $\Psi^{(\rm e)}$ from $k$ and $S$. Fig.~\ref{fig:zerilli} summarizes
  the kind of problem that one can find computing the Zerilli-Moncrief
  function too naively. It refers to $\Psi^{(\rm e)}(r)$ at $t=0$.
  We consider (for simplicity of discussion) a 
  polytropic model, i.e. $p=K \rho^{\Gamma}$ with $K=100$, $\Gamma=2$ 
  and $\rho_c=1.28\times 10^{-3}$ with type 3 initial data (i.e. a
  Gaussian pulse with $\sigma=M$ centered at $r=40$). 
  We fix $\Psi^{(\rm e)}$ by Eq.~\eqref{eq:id:gauss}, 
  we compute $\chi$ and $k$ from Eqs.~\eqref{eq:k}-\eqref{eq:chi},
  we compute $k_{,r}$ numerically and then we 
  we reconstruct $\Psi^{(\rm e)}$ via Eq.~(\ref{eq:ZM}). 
  Fig.~\ref{fig:zerilli} shows that, for low resolution ($J_i=100$),
  using a second-order finite-differencing standard stencil to compute
  $k_{,r}$ is clearly not enough, as the reconstructed (thicker dashed-dot line) 
  and the ``exact'' (solid line) $\Psi^{(\rm e)}$ are very different. 
  Incrasing the resolution (to 500 points) improves the agreement, 
  which is however not perfect yet (thinner dashed-dot line).
  A visible improvement is obtained using higher order finite-differencing
  operators: the figure shows that a 4th-order operator
  (already in the low-resolution case with $J_i=100$ points) is sufficient 
  to have an accurate reconstruction of the Zerilli-Moncrief function.
  The conclusion is that one needs to use {\it at least} 4th-order 
  finite-differencing operators to compute accurately $\Psi^{(\rm e)}$ 
  from $\chi$ and $k$.
  If this is not done, the resulting function is not reliable and 
  it can't be considered a solution of Eq.~\eqref{eq:ZMequation}. 
  Typically, we have seen that, when this kind of inaccuracy is present,
  the amplitude of (part of) the Zerilli-Moncrief function (usually the
  one related to the $w$-mode burst) grows linearly with $r$ instead of
  tending to a constant value for $r\to\infty$.

  We performed extensive simulations to test the code. The scheme is stable
  and permits us to accurately evolve the equations as long as we wish.
  To check convergence of the waves, we run different resolutions 
  and we computed the energy emitted at infinity (last detector) integrating 
  separately the $\l=2$ odd and even contribute of Eq.~\eqref{eq:GWenergy} 
  over the evolution time interval. The value of the energy converges 
  correctly up to second order terms, $\mathcal{O}(\Delta r^2)$.

  To validate the physical results of the code we compare the frequencies 
  extracted from our simulations with values computed via a frequency
  domain approach. In the case of the polytropic EOS, the results 
  reported in Ref.~\cite{Nagar:2004av} showed that the errors on fluid 
  frequencies are less than 1\%. Spacetime frequencies for 
  polytropic EOS models were checked in Ref.~\cite{Bernuzzi:2008rq}. 
  In this case, when the damping times are sufficiently long 
  (i.e. when a narrow Gaussian pulse is used and the star model is very compact),
  it is possibile to estimate $\nu$ and $\tau$ using a fit procedure,
  with an error of the order of 6\%.
  
  The accuracy of the frequencies does not change when we  
  use realistic EOS. To validate this assertion, we compared the 
  frequencies extracted from our waveforms with those of Andersson 
  and Kokkotas~\cite{Andersson:1997rn} for EOS A with 
  mass $M=1.653$, $M=1.447$ and $M=1.050$
  (see Table~A.1 of Ref.~\cite{Andersson:1997rn}). Our results 
  are listed in Table~\ref{tab:checkfrqs}, together with the
  data of~\cite{Andersson:1997rn} for completeness.
  Fluid frequencies are 
  typically captured with an accuracy below 1\%, while spacetime 
  frequencies and damping times can be estimate with decent 
  accuracy (5\%) only for the maximum mass model.
  As a consequence, we expect that a similar accuracy for fluid modes,
  i.e. of the order of 1\%, should be expected for the 47 NS models 
  of Table~\ref{tab:equil}. For completeness, the corresponding 
  frequencies are listed in Table~\ref{tab:frqsA} and Table~\ref{tab:frqsWFF}.

  The simulations we have discussed in the main text of the paper 
  use resolutions of  $J_i=400$ and $J_i=800$ respectively for the 
  odd and even-parity case. The Courant-Friedrichs-Lewy factor is set
  to $\Delta t / \Delta r = 0.4$. For each model, the outer boundary
  of the grid is at $r=600M$. The final evolution time is $t^{\rm end}=3000M$ 
  for even-parity evolutions and $t^{\rm end}=800M$ for the odd-parity ones.

  
  \bibliography{GW_pTOV_rEOS}

\begin{thebibliography}{66}
\expandafter\ifx\csname natexlab\endcsname\relax\def\natexlab#1{#1}\fi
\expandafter\ifx\csname bibnamefont\endcsname\relax
  \def\bibnamefont#1{#1}\fi
\expandafter\ifx\csname bibfnamefont\endcsname\relax
  \def\bibfnamefont#1{#1}\fi
\expandafter\ifx\csname citenamefont\endcsname\relax
  \def\citenamefont#1{#1}\fi
\expandafter\ifx\csname url\endcsname\relax
  \def\url#1{\texttt{#1}}\fi
\expandafter\ifx\csname urlprefix\endcsname\relax\def\urlprefix{URL }\fi
\providecommand{\bibinfo}[2]{#2}
\providecommand{\eprint}[2][]{\url{#2}}

\bibitem[{\citenamefont{Shapiro and Teukolsky}(1983)}]{Shapiro:1983du}
\bibinfo{author}{\bibfnamefont{S.~L.} \bibnamefont{Shapiro}} \bibnamefont{and}
  \bibinfo{author}{\bibfnamefont{S.~A.} \bibnamefont{Teukolsky}},
  \emph{\bibinfo{title}{{Black holes, white dwarfs, and neutron stars: The
  physics of compact objects}}} (\bibinfo{publisher}{Wiley},
  \bibinfo{address}{New York, USA}, \bibinfo{year}{1983}).

\bibitem[{\citenamefont{Kokkotas and Schmidt}(1999)}]{lrr-1999-2}
\bibinfo{author}{\bibfnamefont{K.~D.} \bibnamefont{Kokkotas}} \bibnamefont{and}
  \bibinfo{author}{\bibfnamefont{B.}~\bibnamefont{Schmidt}},
  \bibinfo{journal}{Living Reviews in Relativity} \textbf{\bibinfo{volume}{2}}
  (\bibinfo{year}{1999}),
  \urlprefix\url{http://www.livingreviews.org/lrr-1999-2}.

\bibitem[{\citenamefont{Kokkotas and Ruoff}(2002)}]{Kokkotas:2002ng}
\bibinfo{author}{\bibfnamefont{K.~D.} \bibnamefont{Kokkotas}} \bibnamefont{and}
  \bibinfo{author}{\bibfnamefont{J.}~\bibnamefont{Ruoff}}
  (\bibinfo{year}{2002}), \eprint{gr-qc/0212105}.

\bibitem[{\citenamefont{Andersson and Kokkotas}(1998)}]{Andersson:1997rn}
\bibinfo{author}{\bibfnamefont{N.}~\bibnamefont{Andersson}} \bibnamefont{and}
  \bibinfo{author}{\bibfnamefont{K.~D.} \bibnamefont{Kokkotas}},
  \bibinfo{journal}{Mon. Not. Roy. Astron. Soc.}
  \textbf{\bibinfo{volume}{299}}, \bibinfo{pages}{1059} (\bibinfo{year}{1998}),
  \eprint{gr-qc/9711088}.

\bibitem[{\citenamefont{Baiotti et~al.}(2005)}]{Baiotti:2004wn}
\bibinfo{author}{\bibfnamefont{L.}~\bibnamefont{Baiotti}} \bibnamefont{et~al.},
  \bibinfo{journal}{Phys. Rev.} \textbf{\bibinfo{volume}{D71}},
  \bibinfo{pages}{024035} (\bibinfo{year}{2005}), \eprint{gr-qc/0403029}.

\bibitem[{\citenamefont{Shibata et~al.}(2006)\citenamefont{Shibata, Liu,
  Shapiro, and Stephens}}]{shibata:104026}
\bibinfo{author}{\bibfnamefont{M.}~\bibnamefont{Shibata}},
  \bibinfo{author}{\bibfnamefont{Y.~T.} \bibnamefont{Liu}},
  \bibinfo{author}{\bibfnamefont{S.~L.} \bibnamefont{Shapiro}},
  \bibnamefont{and} \bibinfo{author}{\bibfnamefont{B.~C.}
  \bibnamefont{Stephens}}, \bibinfo{journal}{Phys. Rev. D}
  \textbf{\bibinfo{volume}{74}}, \bibinfo{eid}{104026} (\bibinfo{year}{2006}).

\bibitem[{\citenamefont{Shibata et~al.}(2005)\citenamefont{Shibata, Taniguchi,
  and Ury\={u}}}]{shibata:084021}
\bibinfo{author}{\bibfnamefont{M.}~\bibnamefont{Shibata}},
  \bibinfo{author}{\bibfnamefont{K.}~\bibnamefont{Taniguchi}},
  \bibnamefont{and} \bibinfo{author}{\bibfnamefont{K.}~\bibnamefont{Ury\={u}}},
  \bibinfo{journal}{Phys. Rev. D} \textbf{\bibinfo{volume}{71}},
  \bibinfo{eid}{084021} (\bibinfo{year}{2005}).

\bibitem[{\citenamefont{Dimmelmeier et~al.}(2007)\citenamefont{Dimmelmeier,
  Ott, Janka, Marek, and Muller}}]{dimmelmeier:251101}
\bibinfo{author}{\bibfnamefont{H.}~\bibnamefont{Dimmelmeier}},
  \bibinfo{author}{\bibfnamefont{C.~D.} \bibnamefont{Ott}},
  \bibinfo{author}{\bibfnamefont{H.-T.} \bibnamefont{Janka}},
  \bibinfo{author}{\bibfnamefont{A.}~\bibnamefont{Marek}}, \bibnamefont{and}
  \bibinfo{author}{\bibfnamefont{E.}~\bibnamefont{Muller}},
  \bibinfo{journal}{Phys. Rev. Lett.} \textbf{\bibinfo{volume}{98}},
  \bibinfo{eid}{251101} (\bibinfo{year}{2007}).

\bibitem[{\citenamefont{Dimmelmeier et~al.}(2006)\citenamefont{Dimmelmeier,
  Stergioulas, and Font}}]{Dimmelmeier:2005zk}
\bibinfo{author}{\bibfnamefont{H.}~\bibnamefont{Dimmelmeier}},
  \bibinfo{author}{\bibfnamefont{N.}~\bibnamefont{Stergioulas}},
  \bibnamefont{and} \bibinfo{author}{\bibfnamefont{J.~A.} \bibnamefont{Font}},
  \bibinfo{journal}{Mon. Not. Roy. Astron. Soc.}
  \textbf{\bibinfo{volume}{368}}, \bibinfo{pages}{1609} (\bibinfo{year}{2006}),
  \eprint{astro-ph/0511394}.

\bibitem[{\citenamefont{Kokkotas and Ruoff}(2001)}]{Kokkotas:2000up}
\bibinfo{author}{\bibfnamefont{K.~D.} \bibnamefont{Kokkotas}} \bibnamefont{and}
  \bibinfo{author}{\bibfnamefont{J.}~\bibnamefont{Ruoff}},
  \bibinfo{journal}{Astron. Astrophys.} \textbf{\bibinfo{volume}{366}},
  \bibinfo{pages}{565} (\bibinfo{year}{2001}), \eprint{gr-qc/0011093}.

\bibitem[{\citenamefont{Allen et~al.}(1998)\citenamefont{Allen, Andersson,
  Kokkotas, and Schutz}}]{Allen:1997xj}
\bibinfo{author}{\bibfnamefont{G.}~\bibnamefont{Allen}},
  \bibinfo{author}{\bibfnamefont{N.}~\bibnamefont{Andersson}},
  \bibinfo{author}{\bibfnamefont{K.~D.} \bibnamefont{Kokkotas}},
  \bibnamefont{and} \bibinfo{author}{\bibfnamefont{B.~F.}
  \bibnamefont{Schutz}}, \bibinfo{journal}{Phys. Rev.}
  \textbf{\bibinfo{volume}{D58}}, \bibinfo{pages}{124012}
  (\bibinfo{year}{1998}), \eprint{gr-qc/9704023}.

\bibitem[{\citenamefont{Ruoff}(2001)}]{Ruoff:2001ux}
\bibinfo{author}{\bibfnamefont{J.}~\bibnamefont{Ruoff}},
  \bibinfo{journal}{Phys. Rev.} \textbf{\bibinfo{volume}{D63}},
  \bibinfo{pages}{064018} (\bibinfo{year}{2001}).

\bibitem[{\citenamefont{Nagar et~al.}(2004)\citenamefont{Nagar, Diaz, Pons, and
  Font}}]{Nagar:2004ns}
\bibinfo{author}{\bibfnamefont{A.}~\bibnamefont{Nagar}},
  \bibinfo{author}{\bibfnamefont{G.}~\bibnamefont{Diaz}},
  \bibinfo{author}{\bibfnamefont{J.~A.} \bibnamefont{Pons}}, \bibnamefont{and}
  \bibinfo{author}{\bibfnamefont{J.~A.} \bibnamefont{Font}},
  \bibinfo{journal}{Phys. Rev.} \textbf{\bibinfo{volume}{D69}},
  \bibinfo{pages}{124028} (\bibinfo{year}{2004}), \eprint{gr-qc/0403077}.

\bibitem[{\citenamefont{Andersson and Kokkotas}(1996)}]{Andersson:1996pn}
\bibinfo{author}{\bibfnamefont{N.}~\bibnamefont{Andersson}} \bibnamefont{and}
  \bibinfo{author}{\bibfnamefont{K.~D.} \bibnamefont{Kokkotas}},
  \bibinfo{journal}{Phys. Rev. Lett.} \textbf{\bibinfo{volume}{77}},
  \bibinfo{pages}{4134} (\bibinfo{year}{1996}), \eprint{gr-qc/9610035}.

\bibitem[{\citenamefont{Nagar and Diaz}(2004)}]{Nagar:2004av}
\bibinfo{author}{\bibfnamefont{A.}~\bibnamefont{Nagar}} \bibnamefont{and}
  \bibinfo{author}{\bibfnamefont{G.}~\bibnamefont{Diaz}}, \bibinfo{journal}{in
  {\it Proceedings of 27th Spanish Relativity Meeting (ERE 2003): Gravitational
  Radiation, Alicante, Spain, 11-13 Sep 2003}}  (\bibinfo{year}{2004}),
  \eprint{gr-qc/0408041}.

\bibitem[{\citenamefont{Passamonti et~al.}(2007)\citenamefont{Passamonti,
  Stergioulas, and Nagar}}]{Passamonti:2007tm}
\bibinfo{author}{\bibfnamefont{A.}~\bibnamefont{Passamonti}},
  \bibinfo{author}{\bibfnamefont{N.}~\bibnamefont{Stergioulas}},
  \bibnamefont{and} \bibinfo{author}{\bibfnamefont{A.}~\bibnamefont{Nagar}},
  \bibinfo{journal}{Phys. Rev.} \textbf{\bibinfo{volume}{D75}},
  \bibinfo{pages}{084038} (\bibinfo{year}{2007}), \eprint{gr-qc/0702099}.

\bibitem[{\citenamefont{Passamonti et~al.}(2006)\citenamefont{Passamonti,
  Bruni, Gualtieri, Nagar, and Sopuerta}}]{Passamonti:2005cz}
\bibinfo{author}{\bibfnamefont{A.}~\bibnamefont{Passamonti}},
  \bibinfo{author}{\bibfnamefont{M.}~\bibnamefont{Bruni}},
  \bibinfo{author}{\bibfnamefont{L.}~\bibnamefont{Gualtieri}},
  \bibinfo{author}{\bibfnamefont{A.}~\bibnamefont{Nagar}}, \bibnamefont{and}
  \bibinfo{author}{\bibfnamefont{C.~F.} \bibnamefont{Sopuerta}},
  \bibinfo{journal}{Phys. Rev.} \textbf{\bibinfo{volume}{D73}},
  \bibinfo{pages}{084010} (\bibinfo{year}{2006}), \eprint{gr-qc/0601001}.

\bibitem[{\citenamefont{Ferrari and Kokkotas}(2000)}]{Ferrari:2000sr}
\bibinfo{author}{\bibfnamefont{V.}~\bibnamefont{Ferrari}} \bibnamefont{and}
  \bibinfo{author}{\bibfnamefont{K.~D.} \bibnamefont{Kokkotas}},
  \bibinfo{journal}{Phys. Rev.} \textbf{\bibinfo{volume}{D62}},
  \bibinfo{pages}{107504} (\bibinfo{year}{2000}), \eprint{gr-qc/0008057}.

\bibitem[{\citenamefont{Ruoff et~al.}(2001)\citenamefont{Ruoff, Laguna, and
  Pullin}}]{Ruoff:2000et}
\bibinfo{author}{\bibfnamefont{J.}~\bibnamefont{Ruoff}},
  \bibinfo{author}{\bibfnamefont{P.}~\bibnamefont{Laguna}}, \bibnamefont{and}
  \bibinfo{author}{\bibfnamefont{J.}~\bibnamefont{Pullin}},
  \bibinfo{journal}{Phys. Rev.} \textbf{\bibinfo{volume}{D63}},
  \bibinfo{pages}{064019} (\bibinfo{year}{2001}), \eprint{gr-qc/0005002}.

\bibitem[{\citenamefont{Ferrari
  et~al.}(2004{\natexlab{a}})\citenamefont{Ferrari, Gualtieri, Pons, and
  Stavridis}}]{Ferrari:2003qu}
\bibinfo{author}{\bibfnamefont{V.}~\bibnamefont{Ferrari}},
  \bibinfo{author}{\bibfnamefont{L.}~\bibnamefont{Gualtieri}},
  \bibinfo{author}{\bibfnamefont{J.~A.} \bibnamefont{Pons}}, \bibnamefont{and}
  \bibinfo{author}{\bibfnamefont{A.}~\bibnamefont{Stavridis}},
  \bibinfo{journal}{Mon. Not. Roy. Astron. Soc.}
  \textbf{\bibinfo{volume}{350}}, \bibinfo{pages}{763}
  (\bibinfo{year}{2004}{\natexlab{a}}), \eprint{astro-ph/0310896}.

\bibitem[{\citenamefont{Ferrari
  et~al.}(2004{\natexlab{b}})\citenamefont{Ferrari, Gualtieri, Pons, and
  Stavridis}}]{Ferrari:2004sj}
\bibinfo{author}{\bibfnamefont{V.}~\bibnamefont{Ferrari}},
  \bibinfo{author}{\bibfnamefont{L.}~\bibnamefont{Gualtieri}},
  \bibinfo{author}{\bibfnamefont{J.~A.} \bibnamefont{Pons}}, \bibnamefont{and}
  \bibinfo{author}{\bibfnamefont{A.}~\bibnamefont{Stavridis}},
  \bibinfo{journal}{Class. Quant. Grav.} \textbf{\bibinfo{volume}{21}},
  \bibinfo{pages}{S515} (\bibinfo{year}{2004}{\natexlab{b}}),
  \eprint{astro-ph/0409578}.

\bibitem[{\citenamefont{Stavridis and Kokkotas}(2005)}]{Stavridis:2004hg}
\bibinfo{author}{\bibfnamefont{A.}~\bibnamefont{Stavridis}} \bibnamefont{and}
  \bibinfo{author}{\bibfnamefont{K.~D.} \bibnamefont{Kokkotas}},
  \bibinfo{journal}{Int. J. Mod. Phys.} \textbf{\bibinfo{volume}{D14}},
  \bibinfo{pages}{543} (\bibinfo{year}{2005}), \eprint{gr-qc/0411019}.

\bibitem[{\citenamefont{Nagar}(2004)}]{Nagar:PhD}
\bibinfo{author}{\bibfnamefont{A.}~\bibnamefont{Nagar}}, \bibinfo{journal}{PhD
  Thesis, University of Parma (unpublished)}  (\bibinfo{year}{2004}).

\bibitem[{\citenamefont{Charles W.~{Misner} and
  {Wheeler}}(1973)}]{Gravitation:1973}
\bibinfo{author}{\bibfnamefont{K.~S.~T.} \bibnamefont{Charles W.~{Misner}}}
  \bibnamefont{and} \bibinfo{author}{\bibfnamefont{J.~A.}
  \bibnamefont{{Wheeler}}}, \emph{\bibinfo{title}{Gravitation}}
  (\bibinfo{publisher}{W. H. Freeman and Company}, \bibinfo{address}{San
  Francisco}, \bibinfo{year}{1973}).

\bibitem[{\citenamefont{Gerlach and Sengupta}(1979)}]{Gerlach:1979rw}
\bibinfo{author}{\bibfnamefont{U.~H.} \bibnamefont{Gerlach}} \bibnamefont{and}
  \bibinfo{author}{\bibfnamefont{U.~K.} \bibnamefont{Sengupta}},
  \bibinfo{journal}{Phys. Rev.} \textbf{\bibinfo{volume}{D19}},
  \bibinfo{pages}{2268} (\bibinfo{year}{1979}).

\bibitem[{\citenamefont{Gerlach and Sengupta}(1980)}]{Gerlach:1980tx}
\bibinfo{author}{\bibfnamefont{U.~H.} \bibnamefont{Gerlach}} \bibnamefont{and}
  \bibinfo{author}{\bibfnamefont{U.~K.} \bibnamefont{Sengupta}},
  \bibinfo{journal}{Phys. Rev.} \textbf{\bibinfo{volume}{D22}},
  \bibinfo{pages}{1300} (\bibinfo{year}{1980}).

\bibitem[{\citenamefont{Gundlach and Martin-Garcia}(2000)}]{Gundlach:1999bt}
\bibinfo{author}{\bibfnamefont{C.}~\bibnamefont{Gundlach}} \bibnamefont{and}
  \bibinfo{author}{\bibfnamefont{J.~M.} \bibnamefont{Martin-Garcia}},
  \bibinfo{journal}{Phys. Rev.} \textbf{\bibinfo{volume}{D61}},
  \bibinfo{pages}{084024} (\bibinfo{year}{2000}), \eprint{gr-qc/9906068}.

\bibitem[{\citenamefont{Martin-Garcia and
  Gundlach}(2001)}]{MartinGarcia:2000ze}
\bibinfo{author}{\bibfnamefont{J.~M.} \bibnamefont{Martin-Garcia}}
  \bibnamefont{and} \bibinfo{author}{\bibfnamefont{C.}~\bibnamefont{Gundlach}},
  \bibinfo{journal}{Phys. Rev.} \textbf{\bibinfo{volume}{D64}},
  \bibinfo{pages}{024012} (\bibinfo{year}{2001}), \eprint{gr-qc/0012056}.

\bibitem[{\citenamefont{Regge and Wheeler}(1957)}]{Regge:1957td}
\bibinfo{author}{\bibfnamefont{T.}~\bibnamefont{Regge}} \bibnamefont{and}
  \bibinfo{author}{\bibfnamefont{J.~A.} \bibnamefont{Wheeler}},
  \bibinfo{journal}{Phys. Rev.} \textbf{\bibinfo{volume}{108}},
  \bibinfo{pages}{1063} (\bibinfo{year}{1957}).

\bibitem[{\citenamefont{Zerilli}(1970)}]{Zerilli:1970se}
\bibinfo{author}{\bibfnamefont{F.~J.} \bibnamefont{Zerilli}},
  \bibinfo{journal}{Phys. Rev. Lett.} \textbf{\bibinfo{volume}{24}},
  \bibinfo{pages}{737} (\bibinfo{year}{1970}).

\bibitem[{\citenamefont{Moncrief}(1974)}]{Moncrief:1974am}
\bibinfo{author}{\bibfnamefont{V.}~\bibnamefont{Moncrief}},
  \bibinfo{journal}{Ann. Phys.} \textbf{\bibinfo{volume}{88}},
  \bibinfo{pages}{323} (\bibinfo{year}{1974}).

\bibitem[{\citenamefont{Nagar and Rezzolla}(2005)}]{Nagar:2005ea}
\bibinfo{author}{\bibfnamefont{A.}~\bibnamefont{Nagar}} \bibnamefont{and}
  \bibinfo{author}{\bibfnamefont{L.}~\bibnamefont{Rezzolla}},
  \bibinfo{journal}{Class. Quant. Grav.} \textbf{\bibinfo{volume}{22}},
  \bibinfo{pages}{R167} (\bibinfo{year}{2005}), \eprint{gr-qc/0502064}.

\bibitem[{\citenamefont{Chandrasekhar and
  Ferrari}(1991)}]{Chandrasekhar:1991fi}
\bibinfo{author}{\bibfnamefont{S.}~\bibnamefont{Chandrasekhar}}
  \bibnamefont{and} \bibinfo{author}{\bibfnamefont{V.}~\bibnamefont{Ferrari}},
  \bibinfo{journal}{Proc. Roy. Soc. Lond.} \textbf{\bibinfo{volume}{A432}},
  \bibinfo{pages}{247} (\bibinfo{year}{1991}).

\bibitem[{\citenamefont{Martel and Poisson}(2005)}]{Martel:2005ir}
\bibinfo{author}{\bibfnamefont{K.}~\bibnamefont{Martel}} \bibnamefont{and}
  \bibinfo{author}{\bibfnamefont{E.}~\bibnamefont{Poisson}},
  \bibinfo{journal}{Phys. Rev.} \textbf{\bibinfo{volume}{D71}},
  \bibinfo{pages}{104003} (\bibinfo{year}{2005}), \eprint{gr-qc/0502028}.

\bibitem[{\citenamefont{Pandharipande}(1971{\natexlab{a}})}]{Pandharipande:197%
1:eosA}
\bibinfo{author}{\bibfnamefont{V.~R.} \bibnamefont{Pandharipande}},
  \bibinfo{journal}{Nucl. Phys.} \textbf{\bibinfo{volume}{A174}},
  \bibinfo{pages}{641} (\bibinfo{year}{1971}{\natexlab{a}}).

\bibitem[{\citenamefont{Pandharipande}(1971{\natexlab{b}})}]{Pandharipande:197%
1:eosB}
\bibinfo{author}{\bibfnamefont{V.~R.} \bibnamefont{Pandharipande}},
  \bibinfo{journal}{Nucl. Phys.} \textbf{\bibinfo{volume}{A178}},
  \bibinfo{pages}{123} (\bibinfo{year}{1971}{\natexlab{b}}).

\bibitem[{\citenamefont{Bethe and Johnson}(1974)}]{Bethe:1974:eosC}
\bibinfo{author}{\bibfnamefont{H.~A.} \bibnamefont{Bethe}} \bibnamefont{and}
  \bibinfo{author}{\bibfnamefont{M.~B.} \bibnamefont{Johnson}},
  \bibinfo{journal}{Nucl. Phys.} \textbf{\bibinfo{volume}{A230}},
  \bibinfo{pages}{1} (\bibinfo{year}{1974}).

\bibitem[{\citenamefont{{Lorenz} et~al.}(1993)\citenamefont{{Lorenz},
  {Ravenhall}, and {Pethick}}}]{Lorenz1993:FPS}
\bibinfo{author}{\bibfnamefont{C.~P.} \bibnamefont{{Lorenz}}},
  \bibinfo{author}{\bibfnamefont{D.~G.} \bibnamefont{{Ravenhall}}},
  \bibnamefont{and} \bibinfo{author}{\bibfnamefont{C.~J.}
  \bibnamefont{{Pethick}}}, \bibinfo{journal}{Physical Review Letters}
  \textbf{\bibinfo{volume}{70}}, \bibinfo{pages}{379} (\bibinfo{year}{1993}).

\bibitem[{\citenamefont{{Friedman} and
  {Pandharipande}}(1981)}]{Friedman:1981:FP}
\bibinfo{author}{\bibfnamefont{B.}~\bibnamefont{{Friedman}}} \bibnamefont{and}
  \bibinfo{author}{\bibfnamefont{V.~R.} \bibnamefont{{Pandharipande}}},
  \bibinfo{journal}{Nuclear Physics A} \textbf{\bibinfo{volume}{361}},
  \bibinfo{pages}{502} (\bibinfo{year}{1981}).

\bibitem[{\citenamefont{{Canuto} and {Chitre}}(1974)}]{Canuto:1974:eosG}
\bibinfo{author}{\bibfnamefont{V.}~\bibnamefont{{Canuto}}} \bibnamefont{and}
  \bibinfo{author}{\bibfnamefont{S.~M.} \bibnamefont{{Chitre}}},
  \bibinfo{journal}{\prd} \textbf{\bibinfo{volume}{9}}, \bibinfo{pages}{1587}
  (\bibinfo{year}{1974}).

\bibitem[{\citenamefont{Arnett and Bowers}(1977)}]{Arnett:1976dh}
\bibinfo{author}{\bibfnamefont{W.~D.} \bibnamefont{Arnett}} \bibnamefont{and}
  \bibinfo{author}{\bibfnamefont{R.~L.} \bibnamefont{Bowers}},
  \bibinfo{journal}{Astrophys.J.Suppl.Series} \textbf{\bibinfo{volume}{33}},
  \bibinfo{pages}{415} (\bibinfo{year}{1977}).

\bibitem[{\citenamefont{{Walecka}}(1974)}]{Walecka:1974:eosN}
\bibinfo{author}{\bibfnamefont{J.~D.} \bibnamefont{{Walecka}}},
  \bibinfo{journal}{Annals of Physics} \textbf{\bibinfo{volume}{83}},
  \bibinfo{pages}{491} (\bibinfo{year}{1974}).

\bibitem[{\citenamefont{{Bowers}
  et~al.}(1975{\natexlab{a}})\citenamefont{{Bowers}, {Gleeson}, and {Daryl
  Pedigo}}}]{Bowers:1975:eosO}
\bibinfo{author}{\bibfnamefont{R.~L.} \bibnamefont{{Bowers}}},
  \bibinfo{author}{\bibfnamefont{A.~M.} \bibnamefont{{Gleeson}}},
  \bibnamefont{and} \bibinfo{author}{\bibfnamefont{R.}~\bibnamefont{{Daryl
  Pedigo}}}, \bibinfo{journal}{\prd} \textbf{\bibinfo{volume}{12}},
  \bibinfo{pages}{3043} (\bibinfo{year}{1975}{\natexlab{a}}).

\bibitem[{\citenamefont{{Bowers}
  et~al.}(1975{\natexlab{b}})\citenamefont{{Bowers}, {Gleeson}, and {Daryl
  Pedigo}}}]{Bowers:1975:eosOb}
\bibinfo{author}{\bibfnamefont{R.~L.} \bibnamefont{{Bowers}}},
  \bibinfo{author}{\bibfnamefont{A.~M.} \bibnamefont{{Gleeson}}},
  \bibnamefont{and} \bibinfo{author}{\bibfnamefont{R.}~\bibnamefont{{Daryl
  Pedigo}}}, \bibinfo{journal}{\prd} \textbf{\bibinfo{volume}{12}},
  \bibinfo{pages}{3056} (\bibinfo{year}{1975}{\natexlab{b}}).

\bibitem[{\citenamefont{Douchin and Haensel}(2001)}]{Douchin:2001sv}
\bibinfo{author}{\bibfnamefont{F.}~\bibnamefont{Douchin}} \bibnamefont{and}
  \bibinfo{author}{\bibfnamefont{P.}~\bibnamefont{Haensel}},
  \bibinfo{journal}{Astron. Astrophys.} \textbf{\bibinfo{volume}{380}},
  \bibinfo{pages}{151} (\bibinfo{year}{2001}), \eprint{astro-ph/0111092}.

\bibitem[{\citenamefont{Wiringa et~al.}(1988)\citenamefont{Wiringa, Fiks, and
  Fabrocini}}]{Wiringa:1988tp}
\bibinfo{author}{\bibfnamefont{R.~B.} \bibnamefont{Wiringa}},
  \bibinfo{author}{\bibfnamefont{V.}~\bibnamefont{Fiks}}, \bibnamefont{and}
  \bibinfo{author}{\bibfnamefont{A.}~\bibnamefont{Fabrocini}},
  \bibinfo{journal}{Phys. Rev.} \textbf{\bibinfo{volume}{C38}},
  \bibinfo{pages}{1010} (\bibinfo{year}{1988}).

\bibitem[{\citenamefont{Haensel et~al.}(2007)\citenamefont{Haensel, Potekhin,
  and Yakovlev}}]{Haensel:2007yy}
\bibinfo{author}{\bibfnamefont{P.}~\bibnamefont{Haensel}},
  \bibinfo{author}{\bibfnamefont{A.~Y.} \bibnamefont{Potekhin}},
  \bibnamefont{and} \bibinfo{author}{\bibfnamefont{D.~G.}
  \bibnamefont{Yakovlev}}, \emph{\bibinfo{title}{{Neutron stars 1: Equation of
  state and structure}}} (\bibinfo{publisher}{Springer}, \bibinfo{address}{New
  York, USA}, \bibinfo{year}{2007}).

\bibitem[{\citenamefont{Benhar et~al.}(2004)\citenamefont{Benhar, Ferrari, and
  Gualtieri}}]{Benhar:2004xg}
\bibinfo{author}{\bibfnamefont{O.}~\bibnamefont{Benhar}},
  \bibinfo{author}{\bibfnamefont{V.}~\bibnamefont{Ferrari}}, \bibnamefont{and}
  \bibinfo{author}{\bibfnamefont{L.}~\bibnamefont{Gualtieri}},
  \bibinfo{journal}{Phys. Rev.} \textbf{\bibinfo{volume}{D70}},
  \bibinfo{pages}{124015} (\bibinfo{year}{2004}), \eprint{astro-ph/0407529}.

\bibitem[{\citenamefont{Benhar et~al.}(1999)\citenamefont{Benhar, Berti, and
  Ferrari}}]{Benhar:1998au}
\bibinfo{author}{\bibfnamefont{O.}~\bibnamefont{Benhar}},
  \bibinfo{author}{\bibfnamefont{E.}~\bibnamefont{Berti}}, \bibnamefont{and}
  \bibinfo{author}{\bibfnamefont{V.}~\bibnamefont{Ferrari}},
  \bibinfo{journal}{Mon. Not. Roy. Astron. Soc.}
  \textbf{\bibinfo{volume}{310}}, \bibinfo{pages}{797} (\bibinfo{year}{1999}),
  \eprint{gr-qc/9901037}.

\bibitem[{\citenamefont{{Lindblom} and
  {Detweiler}}(1983)}]{1983ApJS...53...73L}
\bibinfo{author}{\bibfnamefont{L.}~\bibnamefont{{Lindblom}}} \bibnamefont{and}
  \bibinfo{author}{\bibfnamefont{S.~L.} \bibnamefont{{Detweiler}}},
  \bibinfo{journal}{\apj Suppl. Ser.} \textbf{\bibinfo{volume}{53}},
  \bibinfo{pages}{73} (\bibinfo{year}{1983}).

\bibitem[{\citenamefont{Nozawa et~al.}(1998)\citenamefont{Nozawa, Stergioulas,
  Gourgoulhon, and Eriguchi}}]{Nozawa:1998ak}
\bibinfo{author}{\bibfnamefont{T.}~\bibnamefont{Nozawa}},
  \bibinfo{author}{\bibfnamefont{N.}~\bibnamefont{Stergioulas}},
  \bibinfo{author}{\bibfnamefont{E.}~\bibnamefont{Gourgoulhon}},
  \bibnamefont{and} \bibinfo{author}{\bibfnamefont{Y.}~\bibnamefont{Eriguchi}},
  \bibinfo{journal}{Astron. Astrophys. Suppl. Ser.}
  \textbf{\bibinfo{volume}{132}}, \bibinfo{pages}{431} (\bibinfo{year}{1998}),
  \eprint{gr-qc/9804048}.

\bibitem[{\citenamefont{{Salgado} et~al.}(1994)\citenamefont{{Salgado},
  {Bonazzola}, {Gourgoulhon}, and {Haensel}}}]{Salgado:1994:sbghI}
\bibinfo{author}{\bibfnamefont{M.}~\bibnamefont{{Salgado}}},
  \bibinfo{author}{\bibfnamefont{S.}~\bibnamefont{{Bonazzola}}},
  \bibinfo{author}{\bibfnamefont{E.}~\bibnamefont{{Gourgoulhon}}},
  \bibnamefont{and}
  \bibinfo{author}{\bibfnamefont{P.}~\bibnamefont{{Haensel}}},
  \bibinfo{journal}{Astron. Astrophys.} \textbf{\bibinfo{volume}{291}},
  \bibinfo{pages}{155} (\bibinfo{year}{1994}).

\bibitem[{\citenamefont{Stergioulas and Friedman}(1995)}]{Stergioulas:1994ea}
\bibinfo{author}{\bibfnamefont{N.}~\bibnamefont{Stergioulas}} \bibnamefont{and}
  \bibinfo{author}{\bibfnamefont{J.~L.} \bibnamefont{Friedman}},
  \bibinfo{journal}{Astrophys. J.} \textbf{\bibinfo{volume}{444}},
  \bibinfo{pages}{306} (\bibinfo{year}{1995}), \eprint{astro-ph/9411032}.

\bibitem[{\citenamefont{{Baym} et~al.}(1971{\natexlab{a}})\citenamefont{{Baym},
  {Bethe}, and {Pethick}}}]{Baym:1971:BBP}
\bibinfo{author}{\bibfnamefont{G.}~\bibnamefont{{Baym}}},
  \bibinfo{author}{\bibfnamefont{H.~A.} \bibnamefont{{Bethe}}},
  \bibnamefont{and} \bibinfo{author}{\bibfnamefont{C.~J.}
  \bibnamefont{{Pethick}}}, \bibinfo{journal}{Nuclear Physics A}
  \textbf{\bibinfo{volume}{175}}, \bibinfo{pages}{225}
  (\bibinfo{year}{1971}{\natexlab{a}}).

\bibitem[{\citenamefont{Haensel and Pichon}(1994)}]{Haensel:1994:HP94}
\bibinfo{author}{\bibfnamefont{P.}~\bibnamefont{Haensel}} \bibnamefont{and}
  \bibinfo{author}{\bibfnamefont{B.}~\bibnamefont{Pichon}},
  \bibinfo{journal}{Astron. Astrophys.} \textbf{\bibinfo{volume}{283}},
  \bibinfo{pages}{313} (\bibinfo{year}{1994}), \eprint{nucl-th/9310003}.

\bibitem[{\citenamefont{{Baym} et~al.}(1971{\natexlab{b}})\citenamefont{{Baym},
  {Pethick}, and {Sutherland}}}]{Baym:1971:BPS}
\bibinfo{author}{\bibfnamefont{G.}~\bibnamefont{{Baym}}},
  \bibinfo{author}{\bibfnamefont{C.}~\bibnamefont{{Pethick}}},
  \bibnamefont{and}
  \bibinfo{author}{\bibfnamefont{P.}~\bibnamefont{{Sutherland}}},
  \bibinfo{journal}{\apj} \textbf{\bibinfo{volume}{170}}, \bibinfo{pages}{299}
  (\bibinfo{year}{1971}{\natexlab{b}}).

\bibitem[{\citenamefont{{Haensel} and {Proszynski}}(1982)}]{Haensel:1982}
\bibinfo{author}{\bibfnamefont{P.}~\bibnamefont{{Haensel}}} \bibnamefont{and}
  \bibinfo{author}{\bibfnamefont{M.}~\bibnamefont{{Proszynski}}},
  \bibinfo{journal}{\apj} \textbf{\bibinfo{volume}{258}}, \bibinfo{pages}{306}
  (\bibinfo{year}{1982}).

\bibitem[{\citenamefont{Swesty}(1996)}]{Swesty:1996}
\bibinfo{author}{\bibfnamefont{F.~D.} \bibnamefont{Swesty}},
  \bibinfo{journal}{J.Comput.Phys.} \textbf{\bibinfo{volume}{127}},
  \bibinfo{pages}{118} (\bibinfo{year}{1996}), ISSN \bibinfo{issn}{0021-9991}.

\bibitem[{\citenamefont{Davis et~al.}(1972)\citenamefont{Davis, Ruffini, and
  Tiomno}}]{Davis:1972ud}
\bibinfo{author}{\bibfnamefont{M.}~\bibnamefont{Davis}},
  \bibinfo{author}{\bibfnamefont{R.}~\bibnamefont{Ruffini}}, \bibnamefont{and}
  \bibinfo{author}{\bibfnamefont{J.}~\bibnamefont{Tiomno}},
  \bibinfo{journal}{Phys. Rev.} \textbf{\bibinfo{volume}{D5}},
  \bibinfo{pages}{2932} (\bibinfo{year}{1972}).

\bibitem[{\citenamefont{Bernuzzi et~al.}(2008)\citenamefont{Bernuzzi, Nagar,
  and De~Pietri}}]{Bernuzzi:2008rq}
\bibinfo{author}{\bibfnamefont{S.}~\bibnamefont{Bernuzzi}},
  \bibinfo{author}{\bibfnamefont{A.}~\bibnamefont{Nagar}}, \bibnamefont{and}
  \bibinfo{author}{\bibfnamefont{R.}~\bibnamefont{De~Pietri}},
  \bibinfo{journal}{Phys. Rev.} \textbf{\bibinfo{volume}{D77}},
  \bibinfo{pages}{044042} (\bibinfo{year}{2008}), \eprint{arXiv:0801.2090
  [gr-qc]}.

\bibitem[{\citenamefont{Kokkotas and Schutz}(1992)}]{Kokkotas:1992ka}
\bibinfo{author}{\bibfnamefont{K.~D.} \bibnamefont{Kokkotas}} \bibnamefont{and}
  \bibinfo{author}{\bibfnamefont{B.~F.} \bibnamefont{Schutz}},
  \bibinfo{journal}{Mon. Not. Roy. Astron. Soc.}
  \textbf{\bibinfo{volume}{225}}, \bibinfo{pages}{119} (\bibinfo{year}{1992}).

\bibitem[{\citenamefont{{Nollert}}(1999)}]{1999CQGra..16R.159N}
\bibinfo{author}{\bibfnamefont{H.-P.} \bibnamefont{{Nollert}}},
  \bibinfo{journal}{Class. and Q. Grav.} \textbf{\bibinfo{volume}{16}},
  \bibinfo{pages}{159} (\bibinfo{year}{1999}).

\bibitem[{\citenamefont{Ferrari and Gualtieri}(2008)}]{Ferrari:2007dd}
\bibinfo{author}{\bibfnamefont{V.}~\bibnamefont{Ferrari}} \bibnamefont{and}
  \bibinfo{author}{\bibfnamefont{L.}~\bibnamefont{Gualtieri}},
  \bibinfo{journal}{Gen. Rel. Grav.} \textbf{\bibinfo{volume}{40}},
  \bibinfo{pages}{945} (\bibinfo{year}{2008}), \eprint{0709.0657}.

\bibitem[{\citenamefont{Stergioulas and Morsink}(Web Resource)}]{RNS}
\bibinfo{author}{\bibfnamefont{N.}~\bibnamefont{Stergioulas}} \bibnamefont{and}
  \bibinfo{author}{\bibfnamefont{S.}~\bibnamefont{Morsink}},
  \bibinfo{journal}{UWM Centre for Gravitation and Cosmlogy}
  (\bibinfo{year}{Web Resource}),
  \urlprefix\url{http://www.gravity.phys.uwm.edu/rns/}.

\bibitem[{\citenamefont{Haensel and Potekhin}(Web Resource)}]{NSG}
\bibinfo{author}{\bibfnamefont{P.}~\bibnamefont{Haensel}} \bibnamefont{and}
  \bibinfo{author}{\bibfnamefont{A.~Y.} \bibnamefont{Potekhin}},
  \bibinfo{journal}{Neutron Star Group, Ioffe Institute}  (\bibinfo{year}{Web
  Resource}), \urlprefix\url{http://www.ioffe.ru/astro/NSG/NSEOS/}.

\bibitem[{\citenamefont{Haensel and Potekhin}(2004)}]{Haensel:2004nu}
\bibinfo{author}{\bibfnamefont{P.}~\bibnamefont{Haensel}} \bibnamefont{and}
  \bibinfo{author}{\bibfnamefont{A.~Y.} \bibnamefont{Potekhin}},
  \bibinfo{journal}{Astron. Astrophys.} \textbf{\bibinfo{volume}{428}},
  \bibinfo{pages}{191} (\bibinfo{year}{2004}), \eprint{astro-ph/0408324}.

\end{thebibliography}
  
  
\end{document}